\documentclass[lettersize,journal]{IEEEtran}
\usepackage{amsmath,amsfonts}
\usepackage{algorithmic}
\usepackage{algorithm}
\usepackage{array}
\usepackage[caption=false,font=normalsize,labelfont=sf,textfont=sf]{subfig}
\usepackage{textcomp}
\usepackage{stfloats}
\usepackage{url}
\usepackage{verbatim}
\usepackage{graphicx}
\usepackage{cite}
\usepackage{diagbox}
\usepackage{multirow}
\usepackage{lipsum}
\usepackage{amssymb}
\usepackage{booktabs}

\def\m{{\mathbf m}}
\def\s{{\mathbf s}}

\def\0{{\mathbf 0}}
\def\a{{\mathbf a}}

\def\z{{\mathbf z}}
\def\n{{\mathbf n}}
\def\A{{\mathbf A}}
\def\I{{\mathbf I}}
\def\W{{\mathbf W}}

\DeclareMathOperator{\Var}{Var}
\DeclareMathOperator{\prox}{prox}
\DeclareMathOperator{\diag}{diag}
\DeclareMathOperator*{\argmin}{argmin}

\begin{document}

\title{Uncertainty Quantification for Deep Unrolling-Based Computational Imaging}

\author{Canberk Ekmekci,~\IEEEmembership{Student Member,~IEEE}, and Mujdat Cetin,~\IEEEmembership{Fellow,~IEEE}
\thanks{This work was partially supported by the National Science Foundation (NSF) under Grant CCF-1934962.}
\thanks{Canberk Ekmekci is with the Department of Electrical and Computer Engineering, University of Rochester, Rochester, 
NY 14627 USA (e-mail: cekmekci@ur.rochester.edu).}
\thanks{Mujdat Cetin is with the Department of Electrical and Computer Engineering and Goergen Institute for Data Science, University of Rochester, Rochester, NY 14627 USA (e-mail: mujdat.cetin@rochester.edu).}}

\markboth{Journal of \LaTeX\ Class Files,~Vol.~14, No.~8, August~2021}%
{Shell \MakeLowercase{\textit{et al.}}: A Sample Article Using IEEEtran.cls for IEEE Journals}


\maketitle

\begin{abstract}
Deep unrolling is an emerging deep learning-based image reconstruction methodology that bridges the gap between model-based and purely deep learning-based image reconstruction methods. Although deep unrolling methods achieve state-of-the-art performance for imaging problems and allow the incorporation of the observation model into the reconstruction process, they do not provide any uncertainty information about the reconstructed image, which severely limits their use in practice, especially for safety-critical imaging applications. 
In this paper, we propose a learning-based image reconstruction framework that incorporates the observation model into the reconstruction task and that is capable of quantifying epistemic and aleatoric uncertainties, based on deep unrolling and Bayesian neural networks. We demonstrate the uncertainty characterization capability of the proposed framework on magnetic resonance imaging and computed tomography reconstruction problems. We investigate the characteristics of the epistemic and aleatoric uncertainty information provided by the proposed framework to motivate future research on utilizing uncertainty information to develop more accurate, robust, trustworthy, uncertainty-aware, learning-based image reconstruction and analysis methods for imaging problems. We show that the proposed framework can provide uncertainty information while achieving comparable reconstruction performance to state-of-the-art deep unrolling methods.
\end{abstract}


\section{Introduction}
\label{sec:introduction}
\IEEEPARstart{T}{his} article concerns imaging problems where the target image is observed through a linear transformation followed by additive noise. This observation model is quite general and has been used to model a variety of imaging techniques such as computed tomography (CT)~\cite{Elbakri2002CT}, magnetic resonance imaging (MRI)~\cite{Fessler2010MRI}, microscopy~\cite{Sarder2006Microscopy}, and radar imaging~\cite{Potter2010Radar}. 

For this observation model, classical model-based iterative reconstruction methods cast the image reconstruction problem as a regularized least squares problem whose objective function is the sum of a data-fidelity term and a regularizer. The observation model of the imaging problem determines the form of the data-fidelity term, and the prior knowledge about the target image governs the form of the regularizer. After obtaining an analytical expression for the data-fidelity term and choosing a regularization function, such as the total variation (TV) semi-norm~\cite{Chambolle2004TV}, the resulting optimization problem is solved iteratively by using an appropriate iterative optimization algorithm such as alternating direction method of multipliers (ADMM)~\cite{Boyd2011ADMM}, half-quadratic splitting (HQS), and proximal gradient descent (PGD) method~\cite{Parikh2014Proximal}.

Inspired by model-based iterative reconstruction methods and the pioneering work of Gregor and LeCun~\cite{Gregor2010Unfolding} on sparse coding, a deep learning-based image reconstruction methodology, which is often referred to as deep unrolling~\cite{Aggarwal2019MoDL, Adler2018LearnedPrimalDual, Borgerding2017VAMP, Chun2018BCDNet, Diamond2018Unrolled, Liu2019DPUnrolling, Yang2016ADMMNet, Mardani2018NPGD} has emerged to bridge the gap between model-based image reconstruction methods and purely deep neural network-based image reconstruction methods. The common theme among deep unrolling methods is that they often design a deep neural network by replacing some parts of the unrolled iterative reconstruction algorithm with trainable parameters and neural networks. The main advantages of deep unrolling methods are that they explicitly incorporate the observation model into the neural network, hence they enforce data consistency, and the resulting deep neural network is interpretable in the sense that the resulting deep learning-based image reconstruction method is essentially classical model-based reconstruction algorithm with some learnable components. 

\IEEEpubidadjcol

Although deep unrolling methods have the advantage of incorporating domain knowledge and the physics of the imaging problem into the neural network architecture, existing deep unrolling methods do not provide any predictive uncertainty information about the reconstructed image since they rely on non-Bayesian (standard) neural networks to reconstruct the target image from the corrupted measurements. This severely limits their applicability in safety-critical real-world imaging applications such as medical imaging, where uncertainty information is crucial to make accurate decisions. Our perspective is that we can solve this problem by taking a Bayesian approach for uncertainty estimation and using Bayesian neural networks (BNNs)~\cite{Neal1995BayesianNN}. BNNs are probabilistic models that can quantify the inherent uncertainty on the target image for a given measurement vector due to the ill-posed nature of the inverse problem, which is referred to as the \emph{aleatoric} uncertainty~\cite{Kendall2017BayesianNN}, and the uncertainty on the parameters of a neural network, which is referred to as the \emph{epistemic} uncertainty~\cite{Kendall2017BayesianNN}, by putting a probability distribution on the parameters and computing the posterior distribution of the parameters given a training dataset. By using BNNs together with the idea of deep unrolling, we claim that we can provide predictive uncertainty information for the reconstructed image while preserving the advantages of deep unrolling.

The contribution of this paper is three-fold:
\begin{itemize}
	\item By bringing the idea of deep unrolling and Bayesian neural networks together, we propose an uncertainty-quantifying learning-based image reconstruction framework. Our approach characterizes the overall predictive uncertainty, which is composed of aleatoric and epistemic uncertainties. The proposed method first assumes that the aleatoric uncertainty has the form of an additive Gaussian noise, which is implicitly assumed by most state-of-the-art deep unrolled networks as shown in Section \ref{ssec:observations}, and defines a likelihood function such that the aleatoric uncertainty is modeled with a U-shaped neural network~\cite{Ronneberger2015UNet} and, the mean of the likelihood function is represented with a deep unrolled neural network. Then, by following the principles of Bayesian neural networks, the proposed method approximates the posterior distribution of the parameters of the likelihood function using a scalable variational inference method called Monte Carlo (MC) Dropout~\cite{Gal2016MCDropout}. Next, for a given test measurement and a training dataset, the proposed method computes the predictive distribution, which can be later used to obtain the reconstructed image and the epistemic and aleatoric uncertainty maps, via Monte Carlo integration.
	\item We qualitatively evaluate the proposed method on MRI and CT reconstruction problems to validate whether the epistemic and aleatoric uncertainty information provided by the proposed method exhibits the theoretical characteristics of epistemic and aleatoric uncertainties. We show that the epistemic and aleatoric uncertainty maps obtained by the proposed method display some of the key theoretical characteristics of epistemic and aleatoric uncertainties.
	\item To assess the quality of the probabilistic predictions quantitatively, we generate the calibration plots and calculate the calibration metrics of the proposed method for MRI and CT reconstruction problems. We show that due to the modeling assumptions made by the proposed method, the proposed method may output slightly underconfident predictions. We later show that the proposed model can be easily calibrated by using the recalibration method introduced by Kuleshov \emph{et~al.}~\cite{Kulesov2018recalibration} to output calibrated probabilistic predictions.	
\end{itemize}
We show that the proposed framework can achieve comparable reconstruction performance to a state-of-the-art deep unrolling method and provide epistemic and aleatoric uncertainty information about the reconstructed image while incorporating the domain knowledge into the reconstruction process.

\section{Related Work}
\label{sec:relatedwork}
The problem of uncertainty quantification for image reconstruction tasks, e.g.,~\cite{Adler2019CWGAN, Bardsley2012MCMC, Bohm2019VAE, Cai2018aMCMC, Cai2018bMAP, Cochrane2022BNN, Dasgupta2021NF, Edupuganti2021VAE, Ekmekci2021UncertaintyPnP, Ekmekci2021UncertaintyUnfoldingPreliminary, Herrmann2019Bregman, Hoffmann2021BNNensemble, Kitichotkul2021SURE, Liu2019Heteroscedastic, Repetti2019UQCredibleRegions, Schlemper2018UncertaintyMRI, Shang2021BNN, Siahkoohi2020BNN, Sun2020DPI, Tanno2019uncertainty, Tonolini2020VarInf, Xue2019UncertaintyPhaseImaging}, has attracted the attention of the computational imaging community again recently due to recent advancements in deep generative modeling~\cite{BondTaylor2021GenerativeSurvey} and BNNs~\cite{Neal1995BayesianNN, Gal2016MCDropout, Kendall2017BayesianNN}. The state-of-the-art deep learning-based image reconstruction methods performing uncertainty characterization, e.g.,~\cite{Adler2019CWGAN, Bohm2019VAE, Cochrane2022BNN, Dasgupta2021NF, Edupuganti2021VAE, Ekmekci2021UncertaintyPnP, Ekmekci2021UncertaintyUnfoldingPreliminary, Herrmann2019Bregman, Hoffmann2021BNNensemble, Kitichotkul2021SURE, Liu2019Heteroscedastic, Schlemper2018UncertaintyMRI, Shang2021BNN, Siahkoohi2020BNN, Sun2020DPI, Tanno2019uncertainty, Tonolini2020VarInf, Xue2019UncertaintyPhaseImaging}, can be divided into two groups: deep generative model-based reconstruction methods and BNN-based reconstruction methods. 

Deep generative model-based reconstruction methods, e.g.,~\cite{Adler2019CWGAN, Bohm2019VAE, Dasgupta2021NF, Edupuganti2021VAE, Sun2020DPI, Tonolini2020VarInf}, seek to approximate the posterior distribution of the target image with the help of a generative model to characterize the inherent uncertainty in the reconstruction task, i.e., the uncertainty on the target image for a given measurement vector. For example, Adler and Oktem~\cite{Adler2019CWGAN} approximate the posterior distribution of the target image given a measurement vector using a conditional Wasserstein generative adversarial network~\cite{Mirza2014CGAN, ArjovskyWGAN}. Bohm \emph{et~al.}~\cite{Bohm2019VAE} use a variational autoencoder~\cite{Kingma2013auto} to represent the prior distribution of the target image and perform variational inference to learn the true posterior distribution of the latent variable given a measurement vector. Sun and Bouman~\cite{Sun2020DPI} utilize another popular generative model, a flow-based model~\cite{Kobyzev2021Normalizingflows, Papamakarios2021Normalizingflows}, to approximate the posterior distribution of the target image given a measurement vector and adjust the parameters of the flow-based model by minimizing the reverse Kullback-Leibler divergence~\cite{Papamakarios2021Normalizingflows} between the output distribution of the flow-based model and the posterior distribution. After training the generative model, the uncertainty on the target image for a given measurement vector can be quantified by calculating the sample variance of the samples generated from the approximation of the posterior distribution of the target image. 

While deep generative model-based reconstruction methods aim to quantify the inherent uncertainty in the reconstruction task, the goal of Bayesian neural network-based reconstruction methods, e.g.,~\cite{Cochrane2022BNN, Ekmekci2021UncertaintyPnP, Ekmekci2021UncertaintyUnfoldingPreliminary, Hoffmann2021BNNensemble, Schlemper2018UncertaintyMRI, Shang2021BNN, Siahkoohi2020BNN, Tanno2019uncertainty, Xue2019UncertaintyPhaseImaging}, is to quantify either the uncertainty on the parameters of the statistical model or both the inherent uncertainty in the reconstruction task and the uncertainty on the parameters of the statistical model. To the best of the authors' knowledge, Schlemper \emph{et~al.}~\cite{Schlemper2018UncertaintyMRI} presented the first two BNN-based image reconstruction methods for the MRI reconstruction problem, showing the potential of BNNs for uncertainty quantification for imaging problems. Subsequently, many BNN-based image reconstruction methods were developed for various problems such as the neuroimage enhancement~\cite{Tanno2019uncertainty}, phase imaging~\cite{Xue2019UncertaintyPhaseImaging}, seismic imaging~\cite{Siahkoohi2020BNN}, computational optical form measurements~\cite{Hoffmann2021BNNensemble}, single-pixel imaging~\cite{Shang2021BNN}, imaging through scattering media~\cite{Cochrane2022BNN}, and more general image reconstruction problems~\cite{Ekmekci2021UncertaintyPnP, Ekmekci2021UncertaintyUnfoldingPreliminary}. Table \ref{tab:relatedworkcomparison} shows the functional models of and the types of uncertainties quantified by BNN-based image reconstruction methods.

\begin{table}[t]
\centering
\caption{High-level comparison of Bayesian neural network-based image reconstruction methods}
\label{tab:relatedworkcomparison}
\begin{tabular}{lccc}
\toprule
Method & Functional Model & Quantified Uncertainties \\
\midrule
Schlemper \emph{et~al.}~\cite{Schlemper2018UncertaintyMRI} & U-Net~\cite{Ronneberger2015UNet} & Epistemic \& Aleatoric  \\
Schlemper \emph{et~al.}~\cite{Schlemper2018UncertaintyMRI} & DCCNN~\cite{Schlemper2018DCCNN} & Epistemic \& Aleatoric  \\
Tanno \emph{et~al.}~\cite{Tanno2019uncertainty} &  ESPCN~\cite{Shi2016ESCPCN} & Epistemic \& Aleatoric \\
Xue \emph{et~al.}~\cite{Xue2019UncertaintyPhaseImaging} & U-Net~\cite{Ronneberger2015UNet} & Epistemic \& Aleatoric \\
Siahkoohi \emph{et~al.}~\cite{Siahkoohi2020BNN} & DIP~\cite{Lempitsky2018DIP} & Epistemic \\
Hoffmann \emph{et~al.}~\cite{Hoffmann2021BNNensemble} & U-Net~\cite{Ronneberger2015UNet} & Epistemic \\
Shang \emph{et~al.}~\cite{Shang2021BNN} & U-Net~\cite{Ronneberger2015UNet} & Epistemic \& Aleatoric \\
Ekmekci and Cetin~\cite{Ekmekci2021UncertaintyPnP} & DRUNet~\cite{Zhang2021plug} & Epistemic \\
Ekmekci and Cetin~\cite{Ekmekci2021UncertaintyUnfoldingPreliminary}$^*$ & Deep Unrolling & Epistemic \\
Cochrane \emph{et~al.}~\cite{Cochrane2022BNN} & U-Net~\cite{Ronneberger2015UNet} & Epistemic\\
Proposed Framework & Deep Unrolling & Epistemic \& Aleatoric\\
\bottomrule
\multicolumn{3}{l}{$^*$Preliminary version of this work}
\end{tabular}
\end{table}

Table \ref{tab:relatedworkcomparison} highlights the main differences between the proposed framework and the aforementioned BNN-based image reconstruction methods. The main difference between the proposed framework and the methods presented in \cite{Schlemper2018UncertaintyMRI, Tanno2019uncertainty, Xue2019UncertaintyPhaseImaging, Hoffmann2021BNNensemble, Shang2021BNN, Cochrane2022BNN, Siahkoohi2020BNN} is that the proposed framework utilizes the idea of deep unrolling to integrate the observation model into the reconstruction process. Incorporation of physics-based models through data-consistency layers provides some level of interpretability. The DCCNN~\cite{Schlemper2018DCCNN} based method presented in \cite{Schlemper2018UncertaintyMRI} contains data-consistency layers; however, the data-consistency layer in \cite{Schlemper2018UncertaintyMRI} leverages the characteristic properties of the forward operator of the MRI observation model, making it highly specialized for MRI reconstruction. On the other hand, the proposed framework only requires the computation of the adjoint of the forward operator of the observation model, which is a considerably less restrictive requirement. If the forward operator deviates from a Fourier operator, the data consistency layer of the DCCNN-based method requires matrix inversion, which is not computationally feasible for large scale inverse problems.
The difference between the proposed framework and the framework presented in \cite{Ekmekci2021UncertaintyPnP} lies in the difference between end-to-end models and Plug-and-Play (PnP) methods~\cite{Venkatakrishnan2013PnP}. While the BNN-based image reconstruction method presented in \cite{Ekmekci2021UncertaintyPnP} is built upon the idea of Plug-and-Play (PnP) priors~\cite{Venkatakrishnan2013PnP}, which does not require end-to-end training, the proposed framework uses a deep unrolled network as its functional model, which is trained in an end-to-end manner.

We note that the preliminary version of this work appeared in \cite{Ekmekci2021UncertaintyUnfoldingPreliminary} as a conference paper. The work presented in this manuscript extends the preliminary work in \cite{Ekmekci2021UncertaintyUnfoldingPreliminary} in several significant ways. First, \cite{Ekmekci2021UncertaintyUnfoldingPreliminary} involved the quantification of epistemic uncertainty only, whereas this paper proposes both epistemic and aleatoric uncertainty quantification.
Second, unlike \cite{Ekmekci2021UncertaintyUnfoldingPreliminary}, the unrolled neural network in the framework we propose here contains different CNN blocks at each iteration. We have experimentally observed that this change leads to a faster and more stable training stage. Finally, this manuscript contains an extensive set of experiments demonstrating the characteristics of epistemic and aleatoric uncertainties.

\section{Proposed Framework}
\label{sec:proposedframework}
In this section, we present a learning-based image reconstruction framework that can incorporate the observation model into the reconstruction process and quantify epistemic and aleatoric uncertainties arising in imaging problems. We start by introducing the assumed observation model and presenting a probabilistic formulation of deep unrolling methods along with a motivation for bringing in BNNs. This provides the basis for our BNN-based image reconstruction and uncertainty characterization approach, the components of which are described in the rest of this section.

\subsection{Observation Model and the Inverse Problem}
We consider the following observation model.
\begin{equation}
\m = \A \s + \n,
\label{eq:forwardproblem}
\end{equation}
where $\m \in \mathbb{F}^M$ is the measurement vector; $\A \in \mathbb{F}^{M \times N}$ is the forward operator, which is the discrete approximation of the transformation applied by the imaging system; $\s \in \mathbb{F}^{N}$ is the target image; and $\n \sim \mathcal{N}(\0, \sigma_n^2 \I)$ is additive white Gaussian noise, where $\mathbb{F}$ stands for either $\mathbb{R}$ or $\mathbb{C}$. In this section, without loss of generality, we only consider the case where $\mathbb{F} = \mathbb{R}$ since generalizing the proposed framework to cover the case $\mathbb{F} = \mathbb{C}$ is straightforward (see \cite{Ekmekci2021UncertaintyUnfoldingPreliminary} for details). 

For an underdetermined system ($M<N$), the inverse problem, i.e., recovering the target image $\s$ from the measurement vector $\m$, is an ill-posed problem. To narrow down the solution space, we can utilize any prior knowledge about the target image. One way to use such prior knowledge systematically is to treat the inverse problem as a maximum \emph{a posteriori} (MAP) estimation problem, which is defined by
\begin{equation}
\hat{\s} = \argmin_{\s \in \mathbb{R}^{N}} \left\{ \| \A \s - \m \|_2^2 + \beta \psi(\s) \right\},
\label{eq:mapestimationproblem}
\end{equation}
where $\hat{\s}$ is the MAP estimate of the target image, the term $\| \A \s - \m \|_2^2$ is the data-fidelity term, the function $\psi: \mathbb{R}^N \to \mathbb{R}$ is the regularizer that comes from the prior knowledge on the target image, and $\beta > 0$ is the parameter controlling the balance between the data-fidelity term and the regularizer. After deciding on the form of the regularizer, e.g., total variation semi-norm or wavelet transform domain regularization, model-based reconstruction methods solve the problem in \eqref{eq:mapestimationproblem} iteratively by using an appropriate optimization algorithm, e.g., ADMM~\cite{Boyd2011ADMM}, HQS, or PGD~\cite{Parikh2014Proximal}. 

\subsection{Probabilistic Formulation of Deep Unrolling and BNNs}
\label{ssec:observations}
For the inverse problem, which is essentially a regression problem, suppose that the likelihood function $p(\s|\m,\theta)$ has the following form.
\begin{equation}
p(\s | \m, \theta) = \mathcal{N} \left( \s | f_\theta (\m), \sigma^2 \I \right),
\label{eq:gaussianlikelihood}
\end{equation}
where $f_\theta: \mathbb{R}^M \to \mathbb{R}^N$ is a deep unrolled network parametrized by the set of parameters $\theta$, and $\sigma > 0$ is a fixed constant. Assuming that the training dataset $\mathcal{D}$ contains i.i.d.\ pairs of measurement vectors and target images, we can compute a MAP estimate of the set of parameters by solving the following optimization problem.
\begin{equation}
\hat{\theta}_{\text{MAP}} = \argmin_{\theta} \left\{ \frac{1}{2\sigma^2} \sum_{i=1}^{N_\mathcal{D}} \| \s^{[i]} - f_\theta(\m^{[i]}) \|_2^2 - \log p(\theta) \right\}
\label{eq:mapestimateofparameters}
\end{equation}
where $(\m^{[i]}, \s^{[i]})$ is the $i^{\text{th}}$ example in the training dataset, $N_\mathcal{D}$ is the number of examples in the training dataset, and the distribution $p(\theta)$ is the prior distribution of the set of parameters. In the inference stage, for a given measurement vector $\m_*$, we can compute the distribution $p(\s_* | \m_*, \hat{\theta}_{\text{MAP}})$ to make predictions about the target image.

This probabilistic formulation implicitly appears in the training and inference stages of state-of-the-art deep unrolling methods. For instance, if we choose the prior $p(\theta)$ to be standard Gaussian distribution, then finding the MAP estimate of the set of parameters boils down to training the neural network $f_\theta$ using the squared error loss with weight decay, which is a cost function frequently used by deep unrolling methods. In the inference stage, for a given measurement vector $\m_*$, outputting the mean of the distribution $p(\s_* | \m_*,  \hat{\theta}_{\text{MAP}})$ as the reconstructed image is equivalent to feeding the measurement vector $\m_*$ into the trained neural network $f_{ \hat{\theta}_{\text{MAP}}}$. Hence, training and inference procedures followed by many existing deep unrolling methods can be interpreted probabilistically using the formulation above. 

Although such procedures are frequently used to train deep unrolling methods, there are two problems with this approach regarding the characterization of uncertainties. The first problem is that this formulation does not model the uncertainty on the target image for a given measurement vector, i.e., the inherent uncertainty on the reconstruction task, since it assumes that the covariance matrix of the likelihood function is a fixed model parameter. The second problem is that this formulation does not model the uncertainty on the set of parameters because it only uses a point estimate of the set of parameters by following MAP estimation principles.

BNNs~\cite{Neal1995BayesianNN, Jospin2022BNNTutorial, Kendall2017BayesianNN} can solve these two problems. BNNs solve the first problem by defining a likelihood function that models the inherent uncertainty on the reconstruction task. In the case of a Gaussian likelihood, this can be accomplished by representing the covariance matrix of the likelihood function with a neural network. To solve the second problem, BNNs place a prior distribution on the set of parameters of the likelihood function and compute the posterior distribution of the parameters given a training dataset. Then, at the inference stage, BNNs compute the predictive distribution for a given measurement vector $\m_*$ by computing the following integral:
\begin{equation}
p(\s_* | \m_*, \mathcal{D}) = \int p(\s_*|\m_*, \theta) p(\theta | \mathcal{D}) d\theta,
\label{eq:predictivedistribution}
\end{equation}
where the distribution $p(\s_* | \m_*, \mathcal{D})$ is the predictive distribution, and the integral is taken over all possible values of $\theta$. The first term of the integrand, which is the likelihood function, incorporates the inherent uncertainty on the reconstruction task (i.e., aleatoric uncertainty), which is created by the ill-posedness of the inverse problem, into the predictive distribution.
The second term of the integrand, on the other hand, which is the posterior distribution of the parameters, incorporates the uncertainty on the set of parameters (i.e., epistemic uncertainty), which is created by the lack of training examples in the training dataset around the test measurement vector, into the predictive distribution through an integral over the possible parameter values. Thanks to this conceptually simple probabilistic formulation, we can utilize BNNs to quantify both epistemic and aleatoric uncertainties in computational imaging problems.

\subsection{Form of the Likelihood Function}
Based on our observations presented in Section \ref{ssec:observations}, we define the form of likelihood function as follows.
\begin{equation}
p(\s|\m,\theta) = \mathcal{N}(\s| f_\gamma(\m), \diag(\sigma_\delta^2(\m))),
\label{eq:proposedlikelihood}
\end{equation}
where $\theta = \gamma \cup \delta$, and $f_\gamma : \mathbb{R}^M \to \mathbb{R}^N$ and $\sigma_\delta^2: \mathbb{R}^M \to \mathbb{R}^N$ are two neural networks parametrized by sets of parameters $\gamma$ and $\delta$, respectively. The neural network $f_\gamma$ maps a given measurement vector to a point in the target image space, and the neural network $\sigma_\delta^2$ aims to capture the inherent uncertainty on the target image for a given measurement vector.  The form of the likelihood function implicitly assumes that $\s = f_\gamma (\m) + \n_l$, where $\n_l \sim \mathcal{N}(\0, \diag(\sigma_\delta^2(\m)))$, i.e., the aleatoric uncertainty is modeled as additive Gaussian noise. This assumption may be restrictive for severely ill-posed inverse problems since the distribution of the underlying image $\s$ for a given measurement vector $\m$ can be quite multimodal. To circumvent this issue, the likelihood function can be defined as a mixture of Gaussians distribution by following the idea of mixture density networks~\cite[Chapter~5.6]{bishop2006book}; however, this does not scale well to most imaging inverse problems since it requires training several neural networks (in the order of the number of mixture components). Another alternative is to introduce a latent variable for the deep neural network $f$ to model complex uncertainty patterns on the target image, which is referred to as Bayesian neural networks with latent variables~\cite{depeweg2017bnnlv}. However, most deep unrolled networks are not latent variable models, and it is not straightforward to integrate latent variables into the deep unrolled architectures. Due to these reasons, we have decided to model the aleatoric uncertainty as additive Gaussian noise using the neural network $\sigma_\delta^2$.

\begin{figure}[t]
	\centering
	\includegraphics[width=0.95\columnwidth]{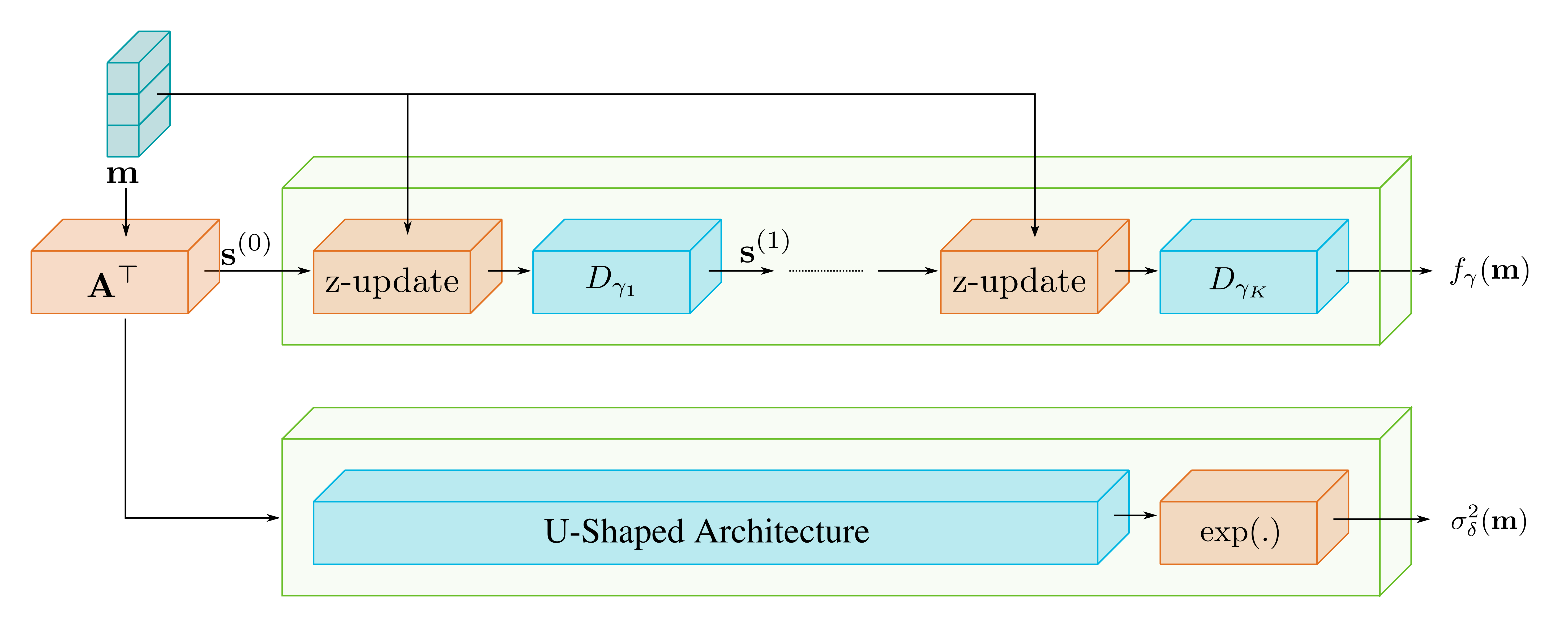}
	\caption{The structure of the neural networks $f_\gamma$ and $\sigma_\delta^2$ at a high level. The neural network $f_\gamma$ maps a measurement vector to a point in the target image space, and the neural network $\sigma_\delta^2$ aims to capture the aleatoric uncertainty. These two neural networks completely specify the form of the Gaussian likelihood in \eqref{eq:proposedlikelihood}.}
	\label{fig:architecture}
\end{figure}

To incorporate the observation model into the neural network $f_\gamma$, which maps a given measurement vector to a point on the target image space, we start constructing $f_\gamma$ by first solving the optimization problem in \eqref{eq:mapestimationproblem} using the proximal gradient descent (PGD) method. The main advantage of using PGD over methods such as ADMM and HQS is that the data dependent update equation of PGD requires computing only the adjoint of the forward operator and does not involve any inversion step, which makes it suitable for large scale imaging problems with non-structured forward operators. Assuming that the regularizer $\psi$ in \eqref{eq:mapestimationproblem} is a closed proper convex function, PGD yields the following iterative image reconstruction algorithm. 
\begin{equation}
\begin{aligned}
\z^{(k+1)} &= (\I - 2\alpha \A^\top \A) \s^{(k)} + 2\alpha \A^\top \m \\
\s^{(k+1)} &= \prox_{\alpha \beta \psi} \left( \z^{(k+1)} \right)
\end{aligned}
\end{equation}
where $\z^{(k+1)} \in\mathbb{R}^N $ is an intermediate vector of the algorithm at the $(k+1)^{\text{st}}$ iteration, $\s^{(k+1)} \in \mathbb{R}^N$ is the reconstructed image at the $(k+1)^{\text{st}}$ iteration, the operator $\prox: \mathbb{R}^N \to \mathbb{R}^N$ is the proximal operator~\cite{Parikh2014Proximal}, and $\alpha > 0$ is the step size. To learn the prior information implicitly from the training data, we replace the proximal operator in the second step with a neural network, which has been frequently done by deep unrolling methods such as \cite{Mardani2018NPGD}. Then, the resulting update equations become
\begin{equation}
\begin{aligned}
\z^{(k+1)} &= (\I - 2\alpha \A^\top \A) \s^{(k)} + 2\alpha \A^\top \m \\
\s^{(k+1)} &= D_{\gamma_{k+1}} \left( \z^{(k+1)} \right),
\end{aligned}
\label{eq:pgdupdatewithnn}
\end{equation}
where $D_{\gamma_{k+1}}: \mathbb{R}^N \to \mathbb{R}^N$ is a residual neural network~\cite{He2016Resnet} parametrized by the set of parameters $\gamma_{k+1}$. For a fixed number of iterations $K$, the series of update equations in \eqref{eq:pgdupdatewithnn} correspond to a deep neural network $f_\gamma$, where $\gamma = \bigcup_{k=1}^K \gamma_k$. Figure \ref{fig:architecture} displays the high-level summary of the neural network $f_\gamma$, and the details of the architecture are provided in the Supplementary Material.

To completely specify the form of the likelihood function given in \eqref{eq:proposedlikelihood}, we have to specify the architecture of the neural network $\sigma_\delta^2$ as well. The neural network we use for the neural network $\sigma_\delta^2$ is a U-shaped neural network~\cite{Ronneberger2015UNet} followed by an element-wise exponentiation to ensure that the output contains positive entries. Figure \ref{fig:architecture} depicts a high-level summary of the neural network $\sigma_\delta^2$, the details of which are given in the Supplementary Material. We must remark that we can also use a dual-head architecture to jointly represent the neural networks $f_\gamma$ and $\sigma_\delta^2$. A brief discussion on the dual-head variant of the proposed framework is also provided in the Supplementary Material for interested readers.

\subsection{Approximating the Posterior Distribution}
\label{ssec:approximatingposterior}
To be able to compute the predictive distribution using \eqref{eq:predictivedistribution}, we have to compute the posterior distribution $p(\theta|\mathcal{D})$. However, exact computation of the posterior distribution is not tractable for deep neural networks because of the massive number of parameters and complex hierarchical structures. Thus, we either have to approximate the posterior distribution with a parametric distribution, or we have to generate samples from the posterior distribution to approximate the integral in \eqref{eq:predictivedistribution}. In our framework, we use a variational inference method called MC Dropout to approximate the posterior distribution with a parametric distribution. The advantages of using MC Dropout are that it is scalable for deep neural networks since it does not introduce additional parameters, variational inference and inference procedures can be straightforwardly implemented in deep learning frameworks because this only requires small changes on the training and testing procedures of the standard neural network pipelines, and it has been shown that MC Dropout provides reliable uncertainty estimates for several problems such as camera relocalization~\cite{Kendall2016CameraRelocalization}, depth completion~\cite{Kendall2017BayesianNN, Gustaffson2020EvaluatingUncertainty} and semantic segmentation~\cite{Kendall2017BayesianNN, Gustaffson2020EvaluatingUncertainty}.

For the sake of completeness, we state the assumptions of MC Dropout explicitly and discuss the variational inference and inference steps. For a more detailed discussion, the reader can refer to ~\cite{Gal2016MCDropout, Gal2016BayesianCNN, Kendall2017BayesianNN}. Suppose that the neural networks $f_\gamma$ and $\sigma_\delta^2$ contain $C$ and $E$ convolutional layers, respectively. Then, we can write the two sets $\gamma$ and $\delta$ as follows:
\begin{equation}
\gamma =  \bigcup_{i=1}^{C} \{ \W_{i}^f \} \quad \text{and} \quad \delta =  \bigcup_{j=1}^E \{ \W_{j}^\sigma \},
\end{equation}
where $\W_{i}^f$ and $\W_j^\sigma$ are the matrices whose rows contain the vectorized filter coefficients of the $i^{\text{th}}$ and $j^{\text{th}}$ convolutional layers of the neural networks $f_\gamma$ and $\sigma^2_{\delta}$, respectively. The assumptions~\cite{Gal2016MCDropout, Gal2016BayesianCNN, Kendall2017BayesianNN} on the parametric distribution $q_\omega (\theta)$ that we use to approximate the true posterior distribution $p(\theta | \mathcal{D})$ are as follows: (i) For the parametric distribution, we assume that the layers of the neural networks $f_\gamma$ and $\sigma_\delta^2$ are independent, and layers within the neural networks are mutually independent, i.e.,
\begin{equation}
q_\omega(\theta) = \left( \prod_{i=1}^{C} q\left(\W_{i}^f\right) \right) \left( \prod_{j=1}^{E} q\left(\W_{j}^\sigma \right) \right);
\end{equation}
(ii) Filters of a convolutional layer are assumed to be mutually independent, more explicitly
\begin{equation}
\begin{aligned}
q(\W_{i}^f) = \prod_{l=1}^{K_{i, f}^{[out]}} q( [ \W_{i}^f ]_{l,:} ), 
q\left(\W_{j}^\sigma \right) = \prod_{m=1}^{K_{j,\sigma}^{[out]}} q( \left[ \W_{j}^\sigma \right]_{m,:} ),
\end{aligned}
\end{equation}
where $K_{i, f}^{[out]}$ is the number of filters in the $i^\text{th}$ convolutional layer of $f_\gamma$, and $K_{j,\sigma}^{[out]}$ is the number of filters in the $j^\text{th}$ convolutional layer of $\sigma^2_\delta$; (iii) The distribution of the filter coefficients of each filter is a mixture of Gaussians distribution defined by
\begin{equation}
\begin{aligned}
q( [ \W_{i}^f ]_{l,:} ) &= p(z_{i,l}^f=1) \mathcal{N}( [ \W_{i}^f ]_{l,:} | \a_{i,l}^f, \epsilon^2\I) \\
&\quad+p(z_{i,l}^f=0) \mathcal{N}( [ \W_{i}^f ]_{l,:} | \0, \epsilon^2\I), \\
q( [ \W_{j}^\sigma ]_{m,:} ) &= p(z_{j,m}^\sigma=1) \mathcal{N}([ \W_{j}^\sigma ]_{m,:} | \a_{j,m}^\sigma, \epsilon^2\I) \\
&\quad+ p(z_{j,m}^\sigma=0) \mathcal{N}([ \W_{j}^\sigma ]_{m,:} | \0, \epsilon^2\I),
\end{aligned}
\label{eq:bernoullivariationaldistribution}
\end{equation}
where the variables $z_{i,l}^f$ and $z_{j,m}^\sigma$ are the latent variables, and the scalars $p_{i,l}^f \triangleq p(z_{i,l}^f=1)$ and $p_{j,m}^\sigma \triangleq p(z_{j,m}^\sigma=1)$ are fixed constants. The scalar $\epsilon$ is a very small fixed constant, and the sets $\Delta_f \triangleq \{ \a_{i,l}^f \}$ and $\Delta_\sigma \triangleq \{ \a_{j,m}^\sigma \}$ are the adjustable parameters of the parametric distribution. Previously we have denoted the set of adjustable parameters of the parametric distribution $q_\omega(\theta)$ with $\omega$, so we can write the set $\omega$ explicitly as $\omega = \Delta_f \cup \Delta_\sigma$.

Under these assumptions, we adjust the parameters of the parametric distribution by minimizing the Kullback-Leibler divergence between the parametric distribution and the true posterior distribution, i.e.,
\begin{equation}
\hat{\omega} = \argmin_\omega D_{\text{KL}} \left( q_\omega(\theta) || p(\theta|\mathcal{D}) \right).
\end{equation}
Under certain approximations and mathematical manipulations (see the Supplementary Material or the Appendix of \cite{Gal2016MCDropout} for the details), the above optimization problem can be approximated with the following optimization problem.
\begin{equation}
\hat{\omega} \approx \argmin_\omega \left\{ g(\omega) + h(\omega)\right\},
\label{eq:variationalinference}
\end{equation}
where 
\begin{equation}
\begin{aligned}
g(\omega) &\triangleq \frac{1}{N_\mathcal{D}} \sum_{n=1}^{N_\mathcal{D}} \sum_{k=1}^{N} \bigg[ \frac{1}{2} \log [\sigma_{\tilde{\delta}^{(n)}}^2 (\m^{[n]})]_k \\
&\mkern-18mu + \frac{1}{2} \exp( -  \log [\sigma_{\tilde{\delta}^{(n)}}^2 (\m^{[n]})]_k) ( [\s^{[n]}]_k - [f_{\tilde{\gamma}^{(n)}} (\m^{[n]})]_k )^2 \bigg], \\
h(\omega) &\triangleq \sum_{i=1}^{C} \sum_{l=1}^{K_{i,f}^{[out]}} \frac{p_{i,l}^f}{2 N_\mathcal{D}} \| \a_{i,l}^f \|_2^2 + \sum_{j=1}^E \sum_{m=1}^{K_{j, \sigma}^{[out]}} \frac{p_{j,m}^\sigma}{2 N_\mathcal{D}} \| \a_{j,m}^\sigma \|_2^2,
\end{aligned}
\label{eq:variationalinferencedefinitions}
\end{equation}
and $\tilde{\theta}^{(n)} = \tilde{\delta}^{(n)} \cup \tilde{\gamma}^{(n)}$ is the $n^{\text{th}}$ sample generated from the parametric distribution $q_\omega(\theta)$. 

After approximating the true posterior distribution $p(\theta | \mathcal{D})$ with the parametric distribution $q_{\hat{\omega}}(\theta)$, we approximate the integral in \eqref{eq:predictivedistribution} using Monte Carlo integration with $T$ samples as follows.
\begin{equation}
\begin{aligned}
p(\s_* | \m_*, \mathcal{D}) &\approx \frac{1}{T} \sum_{t=1}^T  \mathcal{N}(\s_*| f_{\hat{\gamma}^{(t)}}(\m_*), \diag(\sigma^2_{\hat{\delta}^{(t)}}(\m_*)))
\end{aligned}
\label{eq:approximationofpredictivedistribution}
\end{equation}
where $\hat{\theta}^{(t)} = \hat{\delta}^{(t)} \cup \hat{\gamma}^{(t)} $ is the $t^{\text{th}}$ sample from the parametric distribution $q_{\hat{\omega}}(\theta)$. The approximation of the predictive distribution is a mixture of $T$ Gaussians with uniform weights; therefore, we can compute its mean vector and element-wise variance analytically as follows.
\begin{equation}
\mathbb{E}[\s_* | \m_*, \mathcal{D}] \approx \frac{1}{T} \sum_{t=1}^T f_{\hat{\gamma}^{(t)}}(\m_*),
\label{eq:predictivemean}
\end{equation}
\begin{equation}
\begin{aligned}
&\Var[[\s_*]_k | \m_*, \mathcal{D}] \approx \underbrace{\frac{1}{T} \sum_{t=1}^T [\sigma_{\hat{\delta}^{(t)}}^2 (\m_*)]_k}_\text{Aleatoric variance} \\
& + \underbrace{\frac{1}{T} \sum_{t=1}^T [f_{\hat{\gamma}^{(t)}}(\m_*)]_k^2 - \left( \frac{1}{T} \sum_{t=1}^T [f_{\hat{\gamma}^{(t)}}(\m_*)]_k \right)^2}_\text{Epistemic variance},
\end{aligned}
\label{eq:predictivevariance}
\end{equation}
where  $\hat{\theta}^{(t)} = \hat{\delta}^{(t)} \cup \hat{\gamma}^{(t)} $ is the $t^{\text{th}}$ sample from the optimized parametric distribution $q_{\hat{\omega}}(\theta)$. The first term of the predictive variance, which we refer to as the aleatoric variance, reflects the aleatoric uncertainty in the reconstruction task, and the remaining residual sum, which we refer to as the epistemic variance, represents the epistemic uncertainty.

At this point, we have to be aware that we have treated the parameters of the neural networks $f_\gamma$ and $\sigma^2_\delta$ as random variables and have to generate samples from the parametric distribution to solve the optimization problem in \eqref{eq:variationalinference} and to obtain the predictive mean and variance given by \eqref{eq:predictivemean} and \eqref{eq:predictivevariance}. Because we have assumed that filters of convolutional layers are mutually independent, one naive way to generate a sample from the parametric distribution is to generate samples from the distributions in \eqref{eq:bernoullivariationaldistribution}. Sampling from those distributions is equivalent to sampling from a mixture of Gaussians distribution with two components, so, first, we need to sample a Bernoulli random variable, and based on that sample, we generate a sample from one of the two multivariate Gaussian distributions. Because the scalar $\epsilon$ is assumed to be a very small non-zero constant, generating a sample from the multivariate Gaussian distributions in \eqref{eq:bernoullivariationaldistribution} can be approximated by directly reporting the mean. Thus, the whole process of generating a sample from the parametric distribution $q_\omega(\theta)$ boils down to generating samples from Bernoulli random variables and multiplying them with the adjustable parameters of the parametric distribution. Hence we can write
\begin{equation}
\begin{aligned}
\tilde{\gamma}^{(n)} &\approx \left\{ \tilde{z}_{i,l}^{(n)} \a_{i,l}^f | \text{sample }\tilde{z}_{i,l}^{(n)} \sim \text{Bernoulli}(p_{i,l}^f)  \right\}, \\
\tilde{\delta}^{(n)} &\approx \left\{ \tilde{z}_{j,m}^{(n)} \a_{j,m}^\sigma | \text{sample }\tilde{z}_{j,m}^{(n)} \sim \text{Bernoulli}(p_{j,m}^\sigma)  \right\}, \\
\tilde{\theta}^{(n)} &= \tilde{\delta}^{(n)} \cup \tilde{\gamma}^{(n)}.
\end{aligned}
\end{equation}

An interesting observation is that the sampling operation described above resembles the dropout~\cite{Srivastava2014Dropout} operation. Hence, solving the optimization problem in \eqref{eq:variationalinference} boils down to training two neural networks $\bar{f}$ and $\bar{\sigma}^2$ only once using the loss function defined by
\begin{equation}
\begin{aligned}
L(\omega) &= \frac{1}{N_\mathcal{B}} \sum_{n=1}^{N_\mathcal{B}} \sum_{k=1}^{N} \bigg[ \frac{1}{2} \log [\bar{\sigma}_{\Delta_\sigma}^2 (\m^{[n]})]_k \\
&\mkern-18mu + \frac{1}{2} \exp( -  \log [\bar{\sigma}_{\Delta_\sigma}^2 (\m^{[n]})]_k) ( [\s^{[n]}]_k - [\bar{f}_{\Delta_f} (\m^{[n]})]_k )^2 \bigg], \\
\end{aligned}
\label{eq:opt_problem}
\end{equation}
where $\mathcal{B}$ is a mini-batch from the training dataset $\mathcal{D}$, and $N_\mathcal{B}$ is the size of the mini-batch, with weight decay parameters $p_{i,l}^f / (2 N_\mathcal{D})$ and $p_{j,m}^\sigma / (2 N_\mathcal{D})$ and with dropout rates $1-p_{i,l}^f$ and $1-p_{j,m}^\sigma$. After the training stage, the resulting weights of the dropout-added neural networks will be the optimal parameters $\hat{\omega}$ of the parametric distribution $q_{\hat{\omega}}(\theta)$, and generating a sample from the parametric distribution $q_{\hat{\omega}}(\theta)$ simply boils down to applying dropout to the weights of the dropout-added neural networks. Furthermore, calculating the approximation of the predictive distribution using \eqref{eq:approximationofpredictivedistribution} boils down to feeding the test measurement vector to the trained dropout-added neural networks $\bar{f}_{\hat{\Delta}_f}$ and $\bar{\sigma}_{\hat{\Delta}_\sigma}$ exactly $T$ times while the dropout is on. To obtain a reconstruction, we can either generate samples from the approximation of the predictive distribution or compute its mean using \eqref{eq:predictivemean}. To obtain the epistemic and aleatoric uncertainty maps, we use the expression in \eqref{eq:predictivevariance}.

We must remark that we have not included the step size parameter $\alpha$ used in the deep unrolled network $f$ and the parameters of the batch normalization layers of the neural network $\sigma^2$, in the sets $\gamma$ and $\delta$. In other words, we have not taken those parameters into account in the Bayesian formulation. If we want to include the parameters of a batch normalization layer in the formulation, we have to randomly set the parameters of the batch normalization layer to zero, which is not a common practice in deep learning literature. Thus, we decide to treat the parameters of the batch normalization layers as trainable deterministic parameters. Similarly, randomly setting the step size parameter to zero violates the positivity requirement of the step size. We could choose other, preferably more expressive, parametric distributions on the step size parameter to perform variational inference or use deep ensembling~\cite{Lakshminarayanan2017ensembling} to capture the uncertainty on the step size parameter. However, we have experimentally observed that treating the step size parameter and the parameters of the batch normalization layers as trainable deterministic parameters is enough to obtain meaningful uncertainty estimates without introducing more complexity to the model.

\section{Experiments and results}
\label{sec:experiments}
In this section, we present experimental results demonstrating the behavior of our proposed approach. Although the proposed framework can be applied to any inverse problem that can be cast as the optimization problem in \eqref{eq:mapestimationproblem}, we evaluate the proposed framework on basic MRI and CT reconstruction problems as representative applications. We investigate the behavior of epistemic and aleatoric uncertainties under various experimental conditions and show that the epistemic and aleatoric uncertainty information provided by the proposed framework is consistent with the definitions of those uncertainties. We then investigate the calibration properties of the proposed method by generating calibration plots for the MRI and CT reconstruction problems. Finally, we compare the image reconstruction performance of the proposed framework with other image reconstruction methods to demonstrate the image reconstruction capability of the proposed framework. Supplementary Material also contains a toy problem illustrating the concepts of epistemic and aleatoric uncertainties.

\subsection{Experimental Setup}
\label{ssec:experimentalsetup}
\textbf{Datasets:} For the MRI reconstruction problem, we extracted $530$ $256 \times 256$ target images from the IXI Dataset~\cite{ixidataset}. Each target image was normalized between $0$ and $1$. We split the $530$ target images into training, validation, and test datasets containing $500$, $15$, and $15$ target images, respectively. The training, validation, and test datasets were constructed such that they contain target images collected from different subjects. The measurement vectors, i.e., k-space measurements, were generated by computing the subsampled Fourier transform of the target images. For the CT reconstruction problem, we extracted $530$ $512 \times 512$ target images from the LUNA Dataset~\cite{lunadataset}. Each image was resized to $256 \times 256$ pixels and normalized between $0$ and $1$. The training dataset was created by using $500$ target images, and the remaining $30$ images were split into two sets to generate validation and test datasets, each containing $15$ target images. The training, validation, and test datasets were constructed such that they contain target images collected from different subjects. The measurement vectors, i.e., sinogram data, were generated by computing the sparse Radon transform of the target images. Finally, we added white Gaussian noise to the measurement vectors to obtain the final measurement vectors we used in our experiments, where the SNR of the measurement vector is defined as follows:
\begin{equation}
    \text{SNR}(\m_{\text{noiseless}} + \n, \m_{\text{noiseless}}) = 20 \log_{10} \left( \frac{ \| \m_{\text{noiseless}} \|_2}{\|\n\|_2} \right).
\end{equation}

\textbf{Training and Inference Procedures:} Training of the proposed framework refers to solving the optimization problem in \eqref{eq:variationalinference}, where the first term of the objective function is replaced with its mini-batch approximation~\cite{Gal2016MCDropout}. We obtained the neural network $\bar{f}$ by fixing the number of iterations $K$ of the PGD to be $5$ and taking the starting point $\s^{(0)}$ to be the result of zero-filling and filtered backprojection for MRI and CT reconstruction problems, respectively. Each residual block of the neural network $\bar{f}$ contains $5$ convolutional layers, and each convolutional layer is followed by a dropout layer and the leaky ReLU activation function. We used the U-Net architecture for the neural network $\bar{\sigma}^2$, where each convolutional layer is followed by a dropout layer. For the MRI reconstruction problem, the batch size for the training was set to $4$, and the learning rate was fixed to $1\times 10^{-4}$. For the CT reconstruction problem, we used a batch size of $2$ for the training and set the learning rate to $1\times 10^{-5}$. The initial step size $\alpha$ of the PGD algorithm was set to $1.0$ for the MRI experiments and $1\times 10^{-4}$ for the CT experiments. The dropout rate of the dropout layers of the neural networks $\bar{f}$ and $\bar{\sigma}^2$ was set to $0.1$, and the neural networks $\bar{f}$ and $\bar{\sigma}^2$ were trained for $100$ epochs. At the inference stage, a given measurement vector was fed to the neural networks $\bar{f}$ and $\bar{\sigma}^2$ $T=100$ times while the dropout was still activated. The reconstructed image was then obtained by following the approximation in \eqref{eq:predictivemean}. The epistemic and aleatoric uncertainty maps were obtained by calculating three times of the epistemic and aleatoric standard deviations given by \eqref{eq:predictivevariance}.

\subsection{Epistemic Uncertainty}
\label{ssec:epistemicuncertainty}
In theory, epistemic uncertainty is the uncertainty created by the lack of training data around test data and can be explained away by making appropriate changes on the training data. In this subsection, we investigate the characteristics of epistemic uncertainty information provided by the proposed framework and show that the behavior and results of our approach are consistent with the theoretical characteristics of epistemic uncertainty. 

\begin{figure*}[t]
    \centering
    \includegraphics[width=\textwidth]{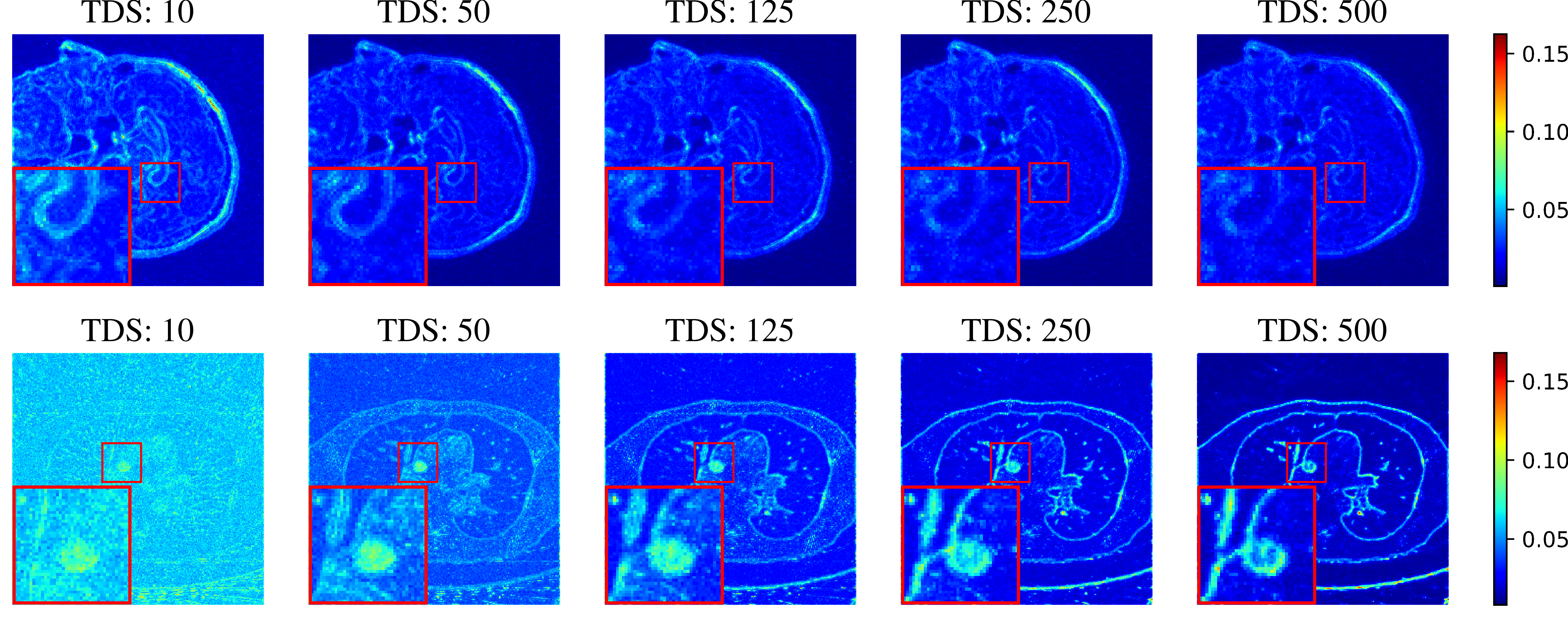}
    \caption{Epistemic uncertainty maps on an MRI (top) and a CT (bottom) test sample as a function of the training dataset size (TDS). As we increase the number of examples in the training dataset, the overall epistemic uncertainty decreases. For the MRI experiments, the percentage of observed k-space coefficients is $20\%$, and SNR is $70$ dB. For the CT experiments, number of views is $36$, and SNR is $70$ dB.}
    \label{fig:reducibilityofepistemicuncertainty}
\end{figure*}

\begin{figure}[t]
    \centering
    \includegraphics[width=0.95\columnwidth]{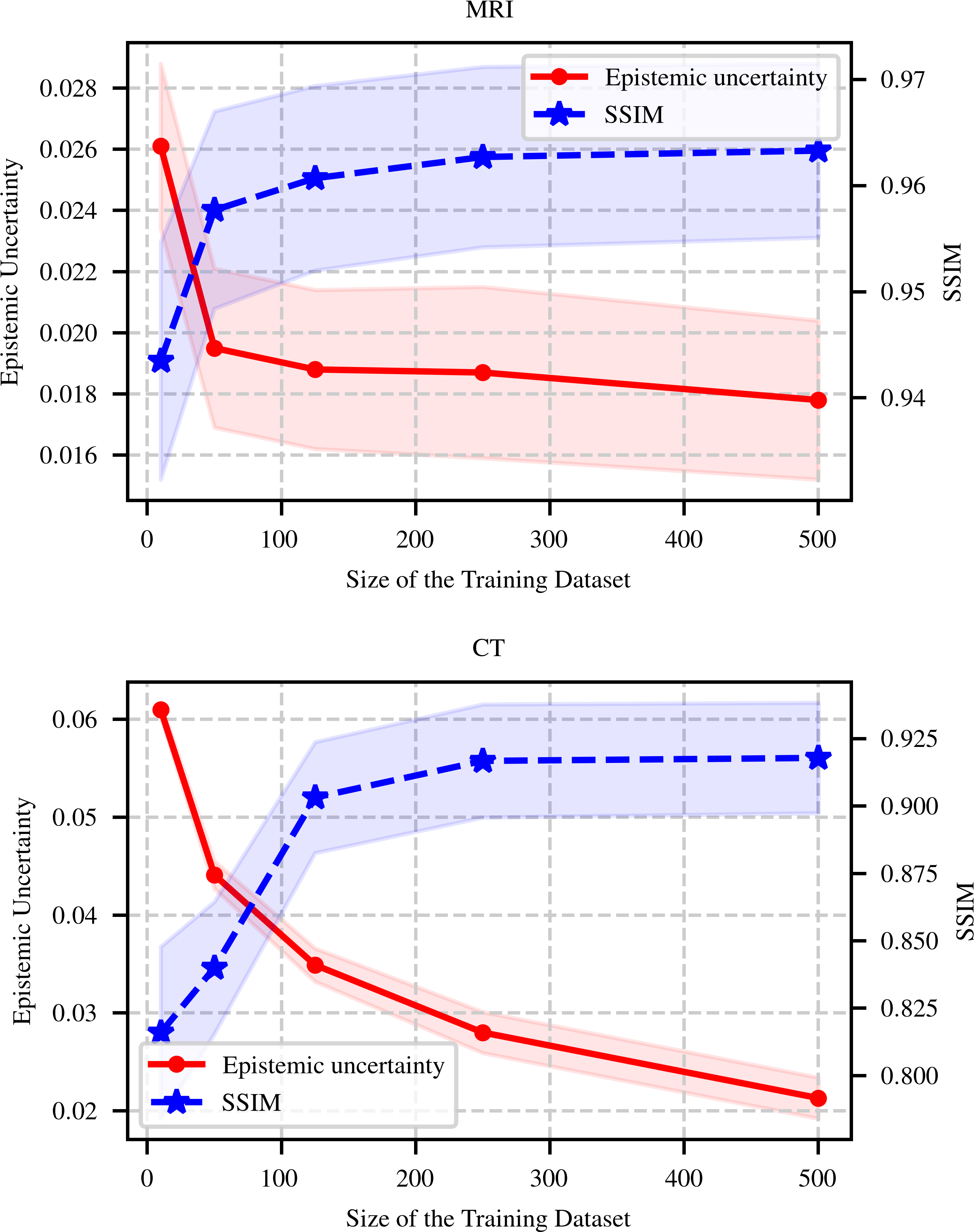}
    \caption{Mean and standard deviation of epistemic uncertainty as a function of training dataset size. The mean and standard deviation are calculated using all pixels in the test dataset. For the MRI experiments, the percentage of observed k-space coefficients is $20\%$, and SNR is $70$ dB. For the CT experiments, number of views is $36$, and SNR is $70$ dB. Mean SSIM values along with the standard deviations for the corresponding reconstructions are provided for reference.}
    \label{fig:reducibilityofepistemicuncertaintyplot}
\end{figure}

To show that the proposed framework outputs epistemic uncertainty information that reflects the uncertainty caused by the lack of training data that can explain the test sample well, we consider two scenarios. In the first scenario, we assess the impact of the size of the training dataset on the inferred epistemic uncertainty. A good uncertainty characterization method should yield larger epistemic uncertainties for smaller training datasets, as it is less probable for such data to represent a random test sample well. In our experiments, we generated five subsets of the MRI training dataset containing $10, 50, 125, 250$, and $500$ examples, and trained five instances of the proposed framework using these subsets as training datasets. Then, for a given test measurement, we obtained the epistemic uncertainty maps using the five trained instances of the proposed framework. We repeated the same procedure for the CT reconstruction problem. The resulting epistemic uncertainty maps are illustrated in Figure \ref{fig:reducibilityofepistemicuncertainty}. For both MRI and CT reconstruction problems, epistemic uncertainty achieves its highest value for the case where we use only $10$ training examples. Then, as we add more examples to the training dataset, epistemic uncertainty on the same test image decreases. To confirm these visual results quantitatively, we calculated the average epistemic uncertainty per pixel taken over the test dataset as a function of the size of the training dataset. Figure \ref{fig:reducibilityofepistemicuncertaintyplot} shows the quantitative results for both MRI and CT reconstruction problems. From this figure, we observe that an increase in the number of training examples leads to a decrease in the overall epistemic uncertainty, which is consistent with the visual results presented in Figure \ref{fig:reducibilityofepistemicuncertainty}. For a discussion about the effect of the size of the training dataset on the aleatoric uncertainty, please refer to the Supplementary Material. 

\begin{figure*}[t]
    \centering
    \includegraphics[width=\textwidth]{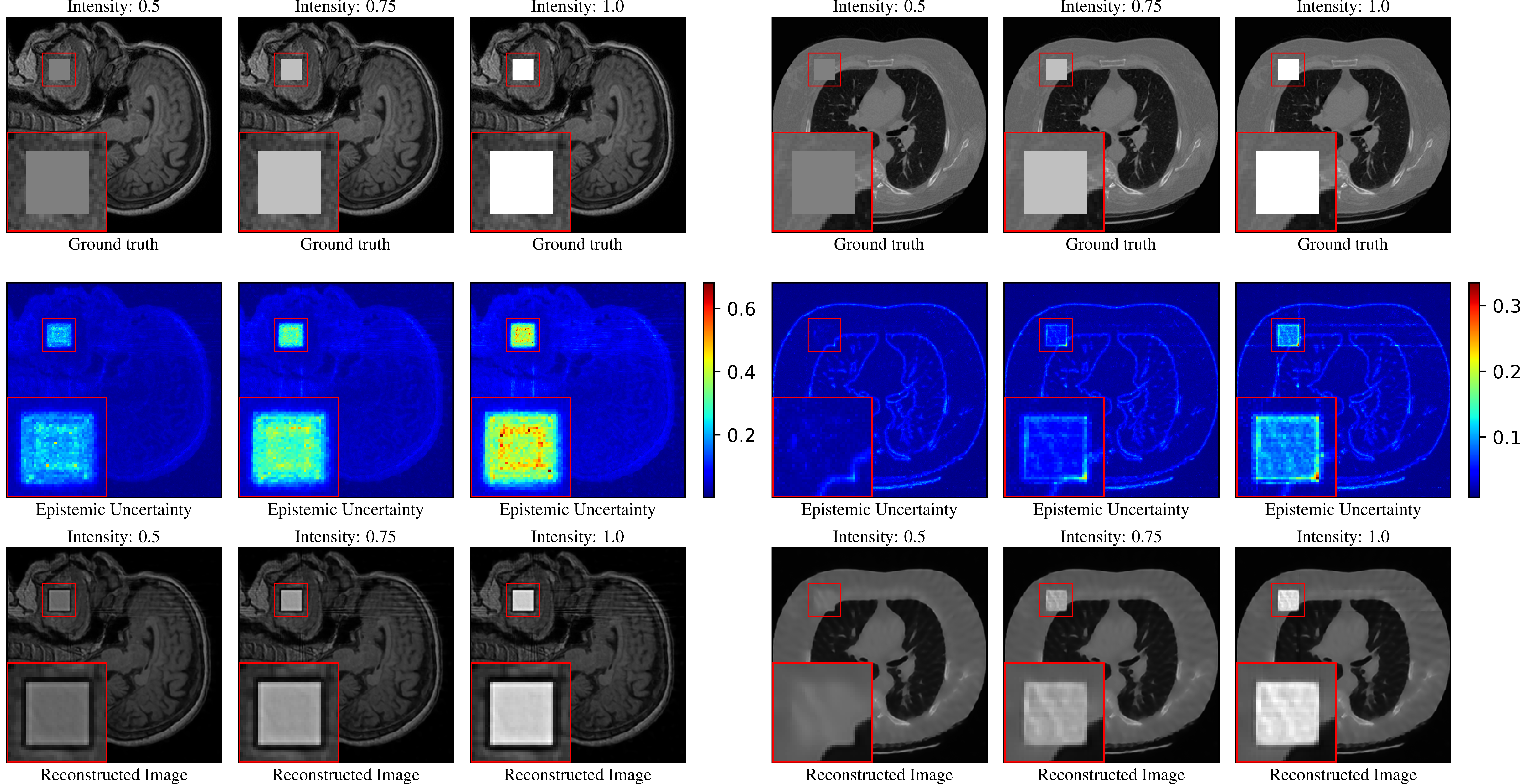}
    \caption{Epistemic uncertainty as a function of the intensity of the abnormal feature. The first row contains the ground truth test images, i.e., the target test images. The second row contains the corresponding epistemic uncertainty maps obtained by the proposed framework, and the third row contains the corresponding reconstructed images. As the inserted square deviates more from the pattern of intensities in the test image (which would be well-represented by the training data), the inferred epistemic uncertainty in the abnormal region increases. For the MRI experiments, the percentage of observed k-space coefficients is $20\%$, and SNR is $70$ dB. For the CT experiments, number of views is $36$, and SNR is $70$ dB.}
    \label{fig:differentintensity}
\end{figure*}

For the second scenario, we insert an artificial feature that is not well-represented by the training dataset to a test target image. Then, we vary the intensity of the inserted abnormal feature to modify the degree of deviation of the test data from the training data. A good uncertainty characterization method would result in larger epistemic uncertainty as the test sample deviates more from the training data. In our experiments, we first trained the proposed framework on the MRI training dataset. Next, we picked a target image from the test dataset and inserted a $25 \times 25$ square with the intensity value of $1.0$ to the test target image. Then, we obtained the epistemic uncertainty map. We repeated the same procedure for different values of the intensity of the inserted abnormal feature and for the CT reconstruction problem. Figure \ref{fig:differentintensity} shows the epistemic uncertainty maps obtained by the proposed framework for different intensity values of the inserted abnormal feature for both MRI and CT reconstruction problems. We observe that the epistemic uncertainty in the abnormal region decreases as the intensity of the inserted square gets close to a value that makes the inserted square visually similar to the target images in the training dataset. Thus, our experiment shows that the epistemic uncertainty map obtained by the proposed framework shows high epistemic uncertainty for test data that are not well-represented by the training data, confirming that for this experiment, the proposed framework successfully captures the uncertainty caused by the lack of training data around the test data.

\begin{figure*}[t]
    \centering
    \includegraphics[width=\textwidth]{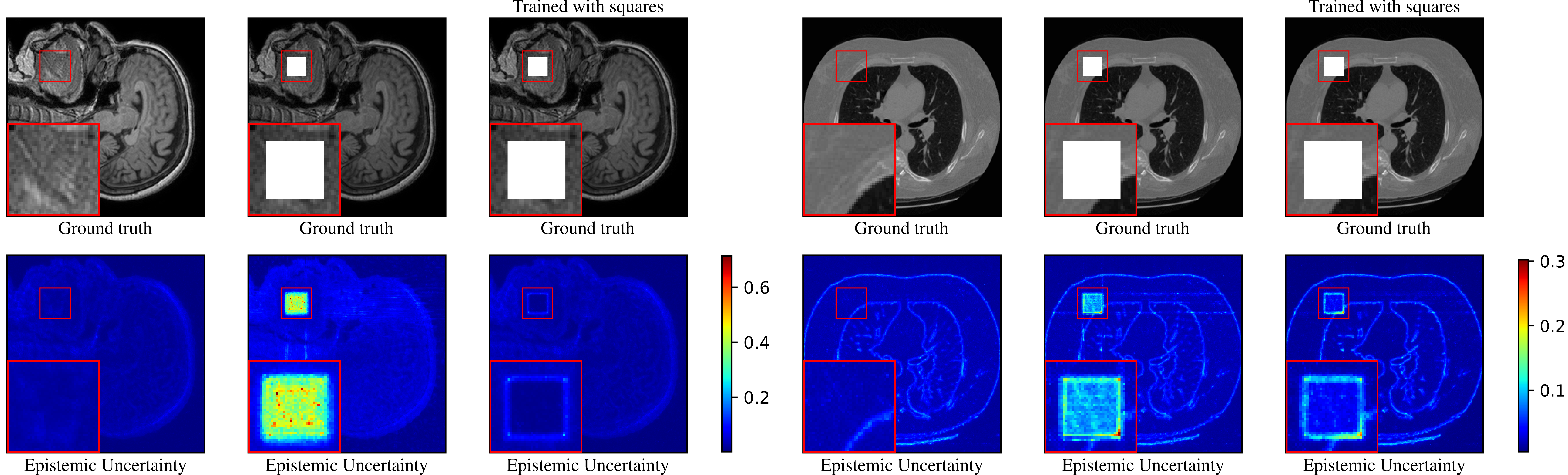}
    \caption{Effect of the structure of the training dataset on epistemic uncertainty maps. The first row contains the ground truth test images, i.e., the target test images, and the second row contains the corresponding epistemic uncertainty maps. The images on the first and fourth columns show the performance of the proposed framework on normal data (i.e., no abnormal features in training and set data). The images on the second and fifth columns show the performance of the proposed framework on a case where an abnormal feature exists in the test data. The images on the third and sixth columns show the performance of the proposed framework with abnormal features present in both training and test data. For the MRI experiments, the percentage of observed k-space coefficients is $20\%$, and SNR is $70$ dB. For the CT experiments, number of views is $36$, and SNR is $70$ dB.}
    \label{fig:outofdataexample}
\end{figure*}

Now, we demonstrate that the epistemic uncertainty provided by the proposed framework possesses the reducibility property. For the first scenario, we have already shown in Figure \ref{fig:reducibilityofepistemicuncertainty} and Figure \ref{fig:reducibilityofepistemicuncertaintyplot} that we can reduce the epistemic uncertainty by collecting more training data having similar characteristics to the test data. For the second scenario, if the proposed framework is capable of capturing the epistemic uncertainty well, we expect to see that using training examples containing features similar to the abnormal feature encountered in the test data would result in reduced epistemic uncertainty. To this end, we added $25 \times 25$ white squares on the training target images and trained the proposed framework with such training data containing the abnormal features. We repeated the same procedure for the CT reconstruction problem. Figure \ref{fig:outofdataexample} shows the resulting epistemic uncertainty maps obtained by the proposed framework for both CT and MRI reconstruction problems. We observe that the epistemic uncertainty around the white square is decreased significantly after we added target images containing white squares into the training dataset, confirming that the epistemic uncertainty provided by the proposed framework can be explained away with additional training data that can represent the test data well. These results qualitatively show that the epistemic uncertainty estimates provided by the proposed method display the reducibility property of the epistemic uncertainty.

\subsection{Aleatoric Uncertainty}
\label{ssec:aleatoricuncertainty}
\begin{figure*}[t]
\centering
\includegraphics[width=0.85\textwidth]{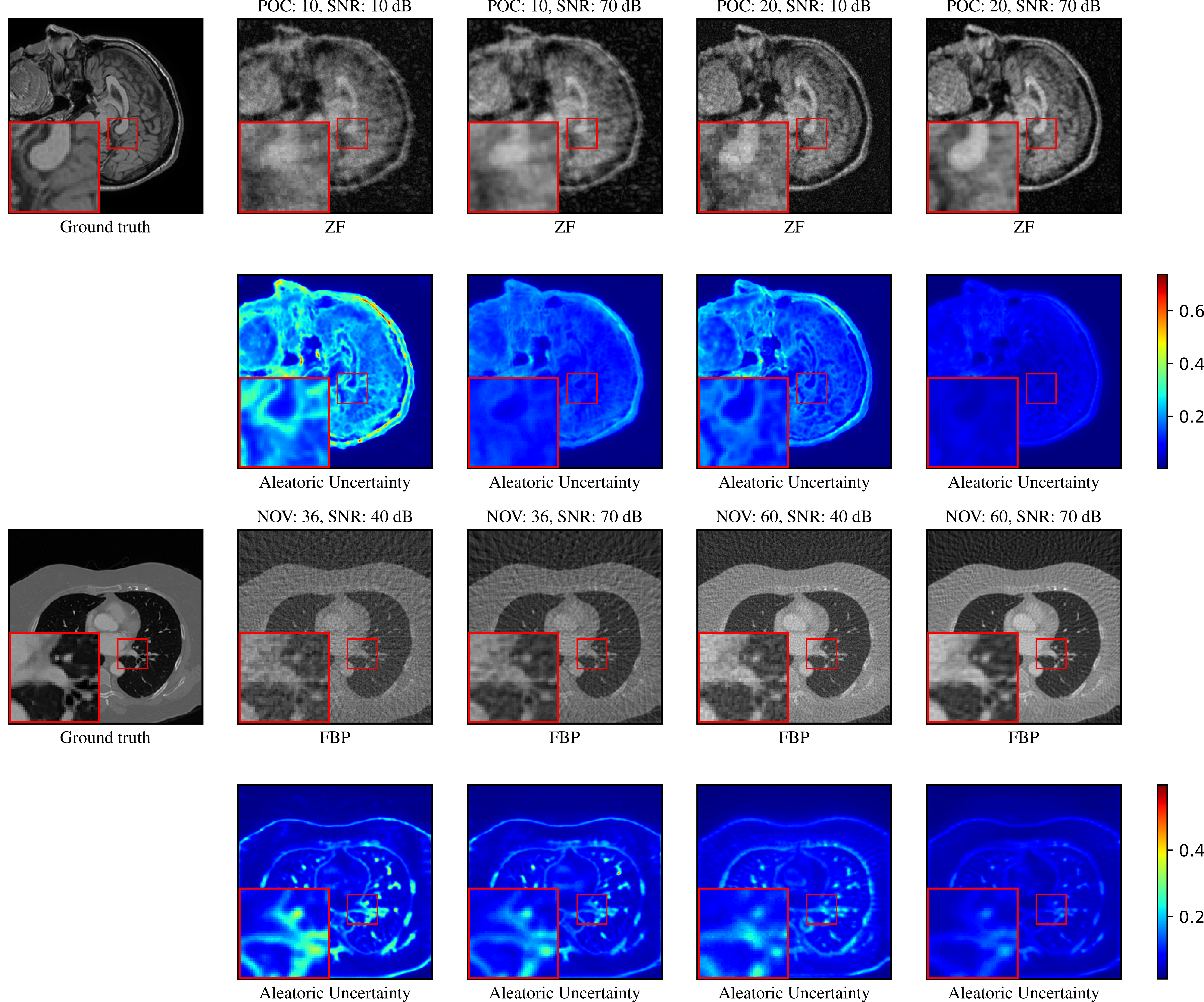}
\caption{Effect of the configuration of the imaging setup on aleatoric uncertainty. The first and the third rows contain the ground truth test images, i.e., target test images, as well as the starting points obtained by applying zero-filling (ZF) or filtered backprojection (FBP) to observations. The second and fourth rows contain the corresponding aleatoric uncertainty maps obtained by the proposed framework for different percentages of observed k-space coefficients (POC), numbers of views (NOV), and signal-to-noise ratios (SNR). Regions where the reconstruction from the starting point is challenging are the regions for which the aleatoric uncertainty is high. Moreover, the overall aleatoric uncertainty increases as the reconstruction problem gets more challenging in terms of data quality and quantity limitations.}
\label{fig:aleatoricuncertaintyandchallengingness}
\end{figure*}

We now focus on aleatoric uncertainty characterization using our proposed framework. The experiments we present here demonstrate successful aleatoric uncertainty characterization, in particular, the aleatoric uncertainty captured by the proposed framework is high for the regions where the reconstruction is challenging due to the ill-posed nature of the inverse problem. Furthermore, we show that the overall aleatoric uncertainty provided by the proposed framework is an indication of how challenging the inverse problem is. For this analysis, we trained the proposed framework for various configurations of the imaging setups. We considered different percentages of observed k-space coefficients and SNR values for the MRI reconstruction problem and different number of views and SNR values for the CT reconstruction problem. Figure \ref{fig:aleatoricuncertaintyandchallengingness} shows the starting points of the proposed framework, i.e., the results of zero-filling and filtered backprojection, and the aleatoric uncertainty maps for different test measurement vectors generated from the two test target images using different configurations of the MR and CT imaging setups.

\begin{figure}[t]
    \centering
    \includegraphics[width=0.95\columnwidth]{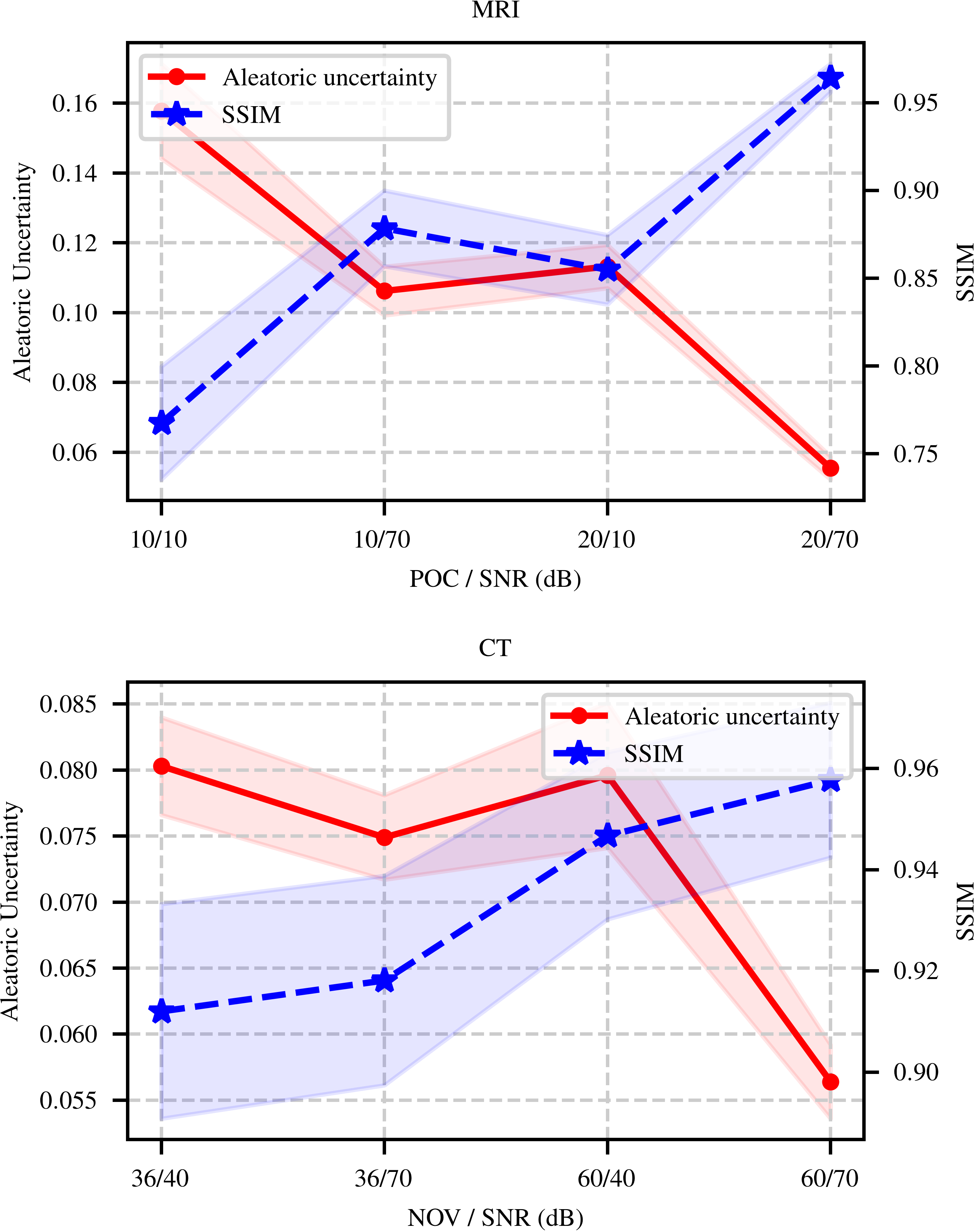}
    \caption{Mean and standard deviation of aleatoric uncertainty for different configurations of the imaging setups. For MRI experiments, they are calculated for different percentages of observed k-space coefficients (POC) and signal-to-noise ratios (SNR). For CT experiments, they are calculated for different numbers of views (NOV) and signal-to-noise ratios (SNR). The averages and standard deviations are calculated using all pixels in the test dataset. Mean SSIM values along with the standard deviations for the corresponding reconstructions are provided for reference.}
    \label{fig:aleatoricuncertaintyandchallengingnessplot}
\end{figure}

For both MRI and CT reconstruction problems, we observe that the aleatoric uncertainty is high for the regions where the reconstruction is challenging for the unrolled network, such as the small localized structures and thin edges on the target images. On the other hand, we observe that the aleatoric uncertainty is low around the regions where the corruption is negligible or can be recovered using the spatial information, such as the smooth regions in the target images. This behavior can be understood analytically with a careful inspection of the objective function of the optimization problem in \eqref{eq:variationalinference}. To minimize the objective function of the optimization problem given in \eqref{eq:variationalinference}, the optimization algorithm needs to minimize the term $\log [\bar{\sigma}^2 (\m^{[n]})]_k$ for the $k^\text{th}$ pixel. However, the term $\exp( -  \log [\bar{\sigma}^2 (\m^{[n]})]_k)$ would increase exponentially if the aforementioned term is minimized. Hence, the value of the term $\log [\bar{\sigma}^2 (\m^{[n]})]_k$ could be made small by the optimization algorithm if the squared error between the output of the neural network $\bar{f}$ and the target image, i.e., $( [\s^{[n]}]_k - [\bar{f}(\m^{[n]})]_k )^2$, is small. In other words, solving optimization problem in \eqref{eq:variationalinference}, which corresponds to performing variational inference using MC Dropout, explicitly forces the dropout added neural network $\bar{\sigma}^2$ to output lower values where the reconstruction is relatively easy. On the other hand, for the regions where the reconstruction is challenging, i.e., for the regions where the squared error between the output of the neural network $\bar{f}$ and the target image is high, solving this optimization problem forces the neural network $\bar{\sigma}^2$ to output high values. Moreover, we observe that the overall aleatoric uncertainty levels show a decrease as SNR decreases for a fixed percentage of the observed k-space coefficients/number of views. Similarly, for a fixed value of the SNR, we observe a decrease in the overall aleatoric uncertainty levels as the percentage of the observed k-space coefficients/number of views increases. Figure \ref{fig:aleatoricuncertaintyandchallengingnessplot} shows the average aleatoric uncertainty over all pixels in the test dataset for different configurations of the imaging setups. From this figure, we observe that the overall aleatoric uncertainty increases when the SNR decreases for a fixed percentage of the observed k-space coefficients/number of views or when the percentage of the observed k-space coefficients/number of views decreases for a fixed value of the SNR. Hence, the quantitative results shown in Figure \ref{fig:aleatoricuncertaintyandchallengingnessplot} confirm our visual observations about the overall aleatoric uncertainty. This result can be also understood by analyzing the objective function of the optimization problem in \eqref{eq:variationalinference}. Because the neural network $\bar{f}$ does not have an infinite learning capability in practice, we expect that the squared error between the output of the trained neural network $\bar{f}$ and the target image will increase as the reconstruction problem gets more challenging, leading to higher overall aleatoric uncertainty levels for the relatively more challenging image reconstruction problems.

\subsection{Calibration Plots of the Proposed Method}
So far, we have observed that epistemic and aleatoric uncertainty maps convey useful information about the confidence of the reconstruction method and the imaging problem; however, we need to perform a more quantitative analysis to evaluate the probabilistic predictions of the proposed framework more reliably. One way of assessing the accuracy of the probabilistic predictions is to look at the calibration and the sharpness properties of the proposed model. In this subsection, we present calibration plots of the proposed method for the MRI and CT reconstruction problems. Furthermore, we briefly touch on the recalibration of the proposed method to achieve more calibrated probabilistic predictions. Since the sharpness metric is useful to compare two probabilistic models, we have included the discussion of the sharpness of the proposed method in the supplementary material, where we compare the probabilistic predictions of the uncertainty-quantifying PnP method presented in \cite{Ekmekci2021UncertaintyPnP} with the proposed method.

\begin{figure}[!t]
    \subfloat[$10\%$, $10$]{%
      \includegraphics[width=0.45\columnwidth]{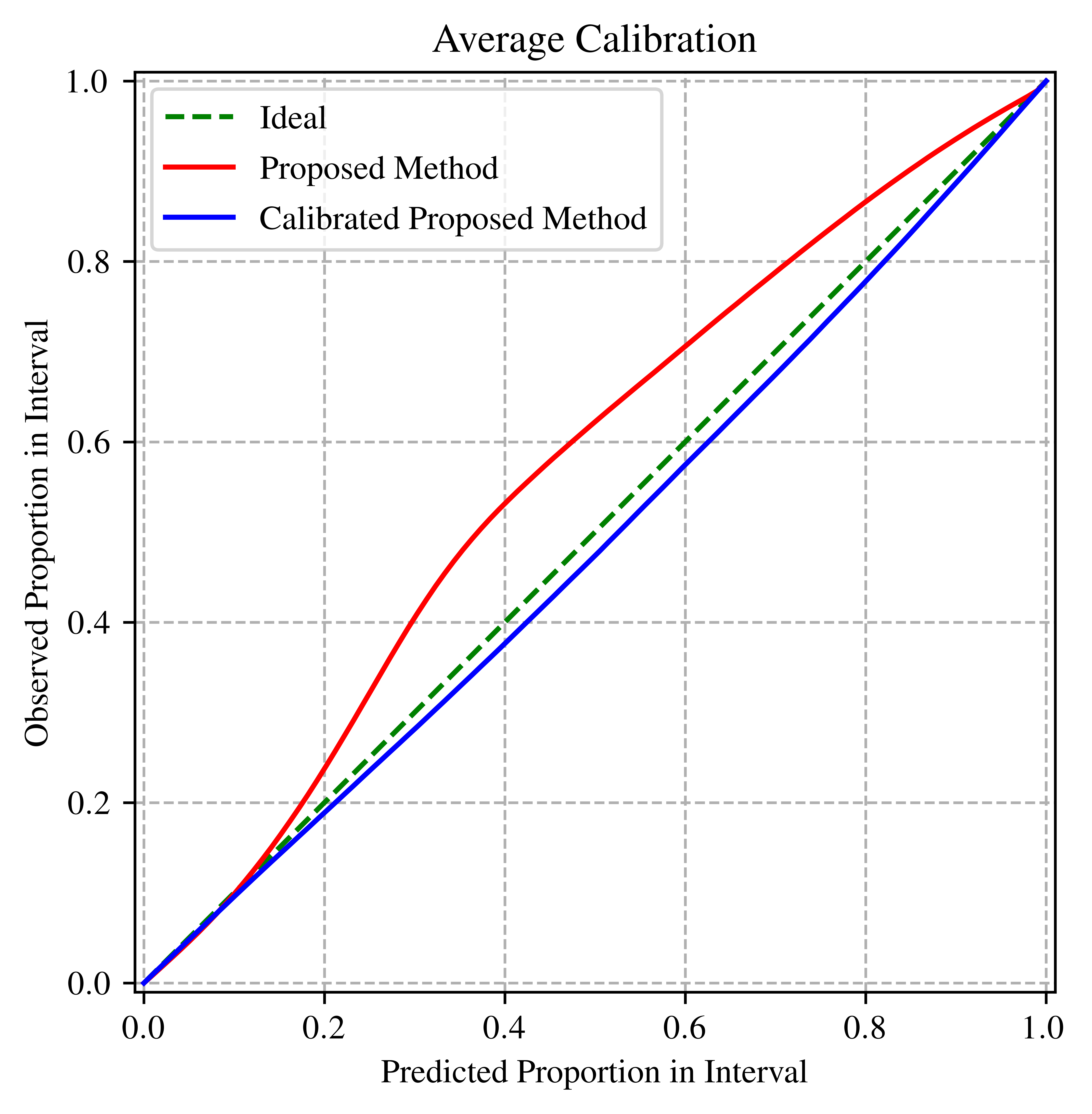}
    }
    \hfill
   \subfloat[$10\%$, $70$]{%
      \includegraphics[width=0.45\columnwidth]{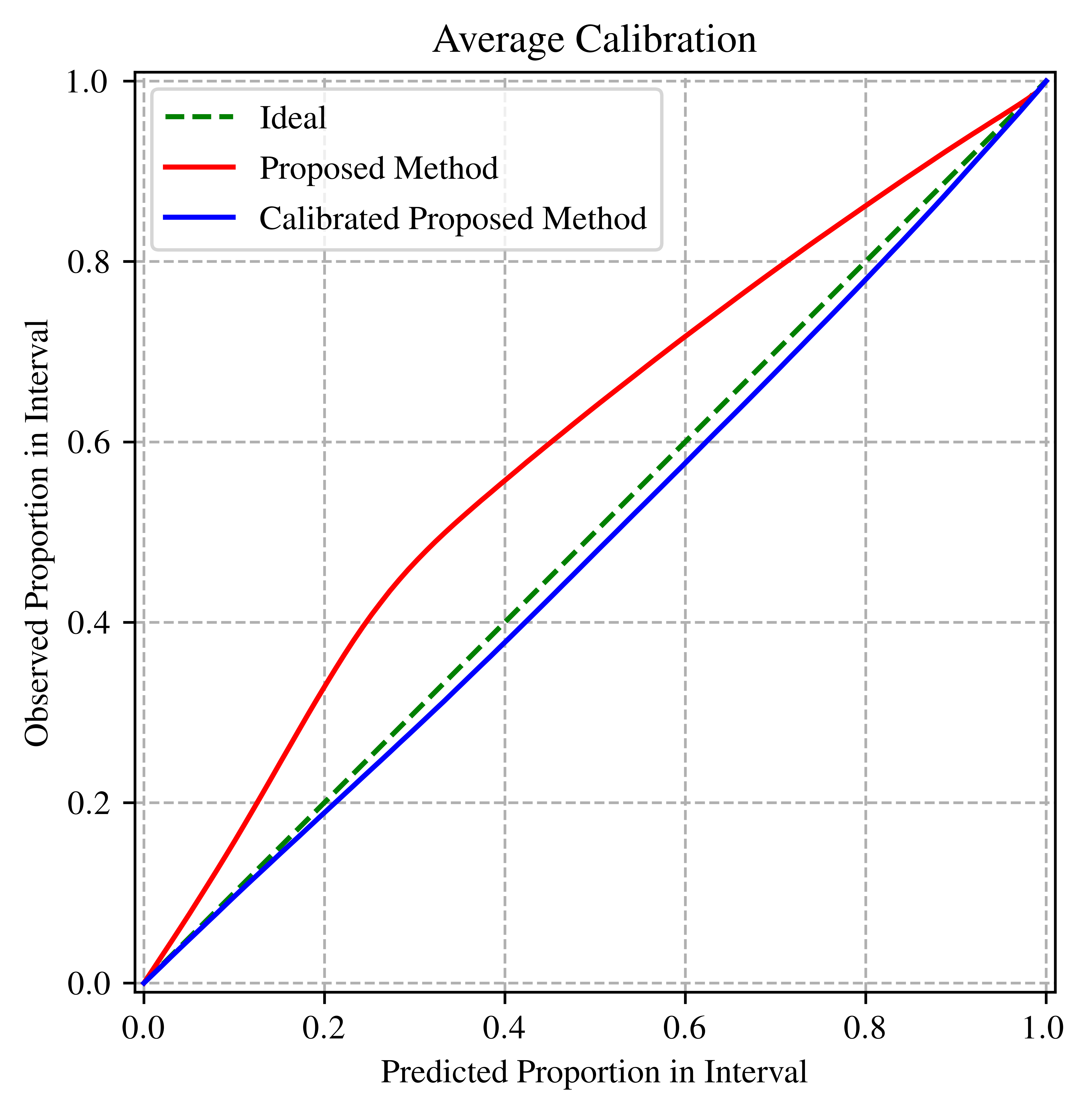}
    } \\
    \subfloat[$20\%$, $10$]{%
      \includegraphics[width=0.45\columnwidth]{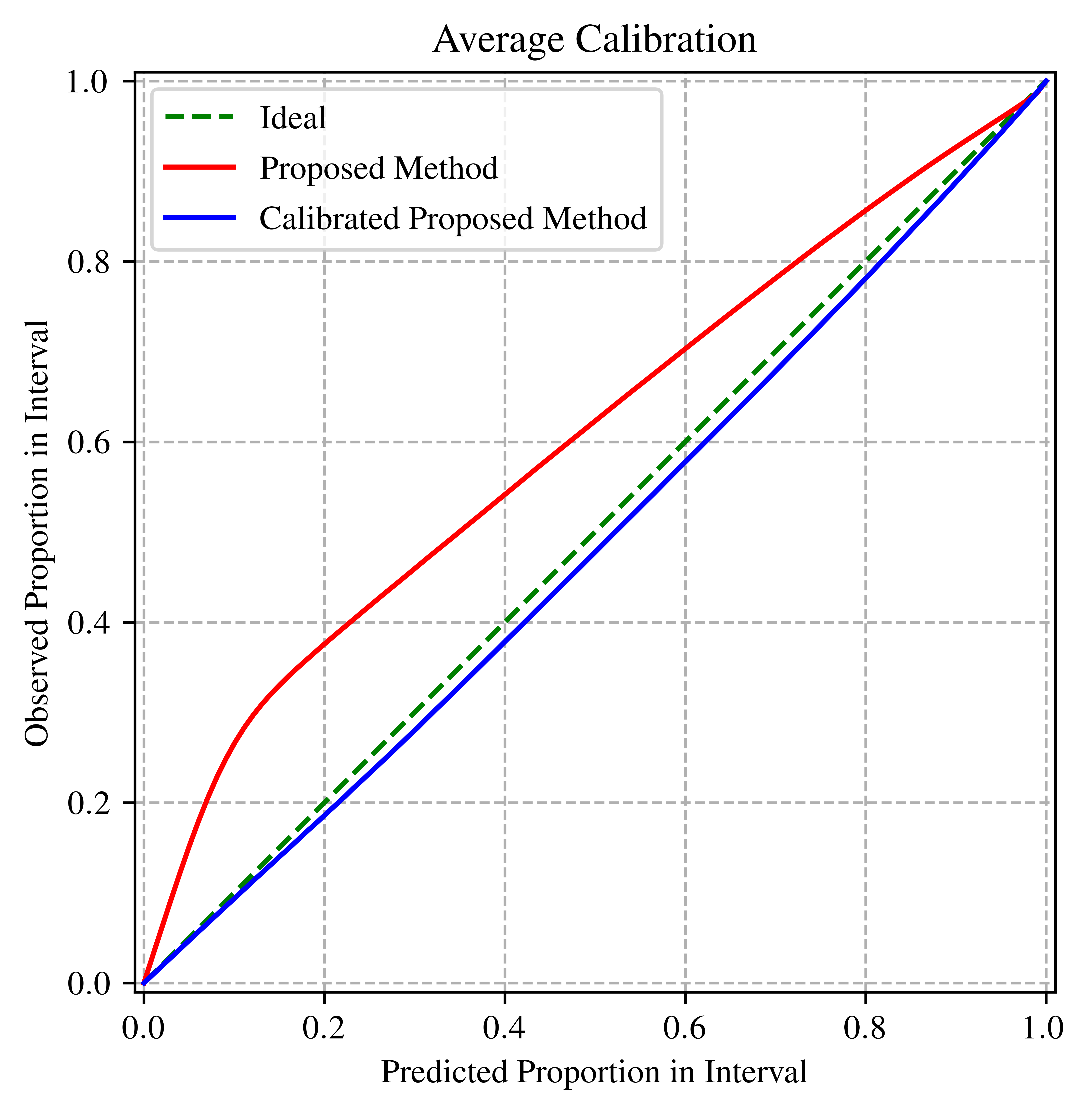}
    }
    \hfill
    \subfloat[$20\%$, $70$]{%
      \includegraphics[width=0.45\columnwidth]{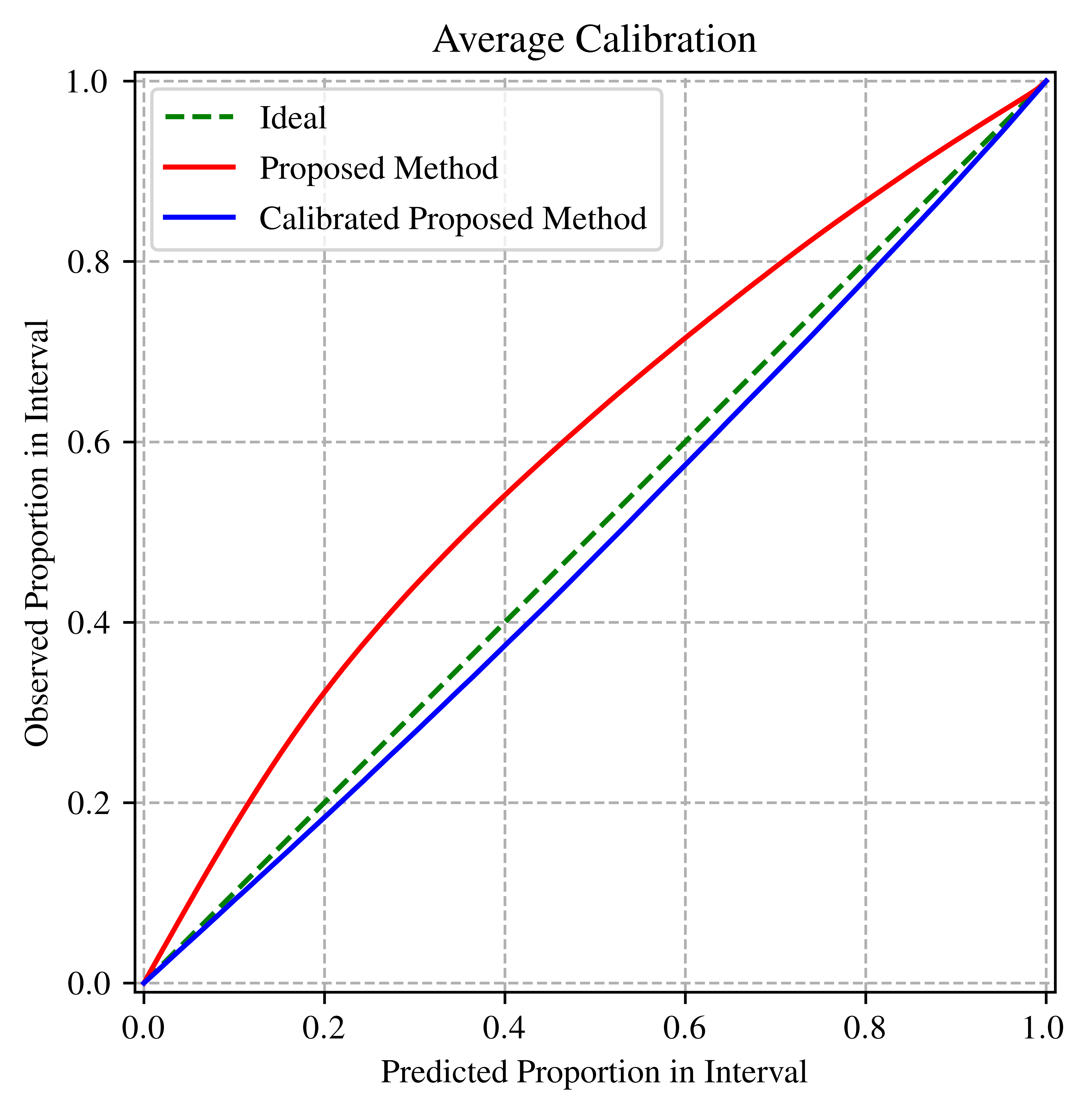}
    }
    \caption{Calibration plots of the proposed method for different configurations of the MRI setup. Subcaptions specify the percentage of observed k-space coefficients and the SNR (dB).}
    \label{fig:calibration_plots_mri}
\end{figure}

\begin{figure}[!t]
    \subfloat[$36$, $40$]{%
      \includegraphics[width=0.45\columnwidth]{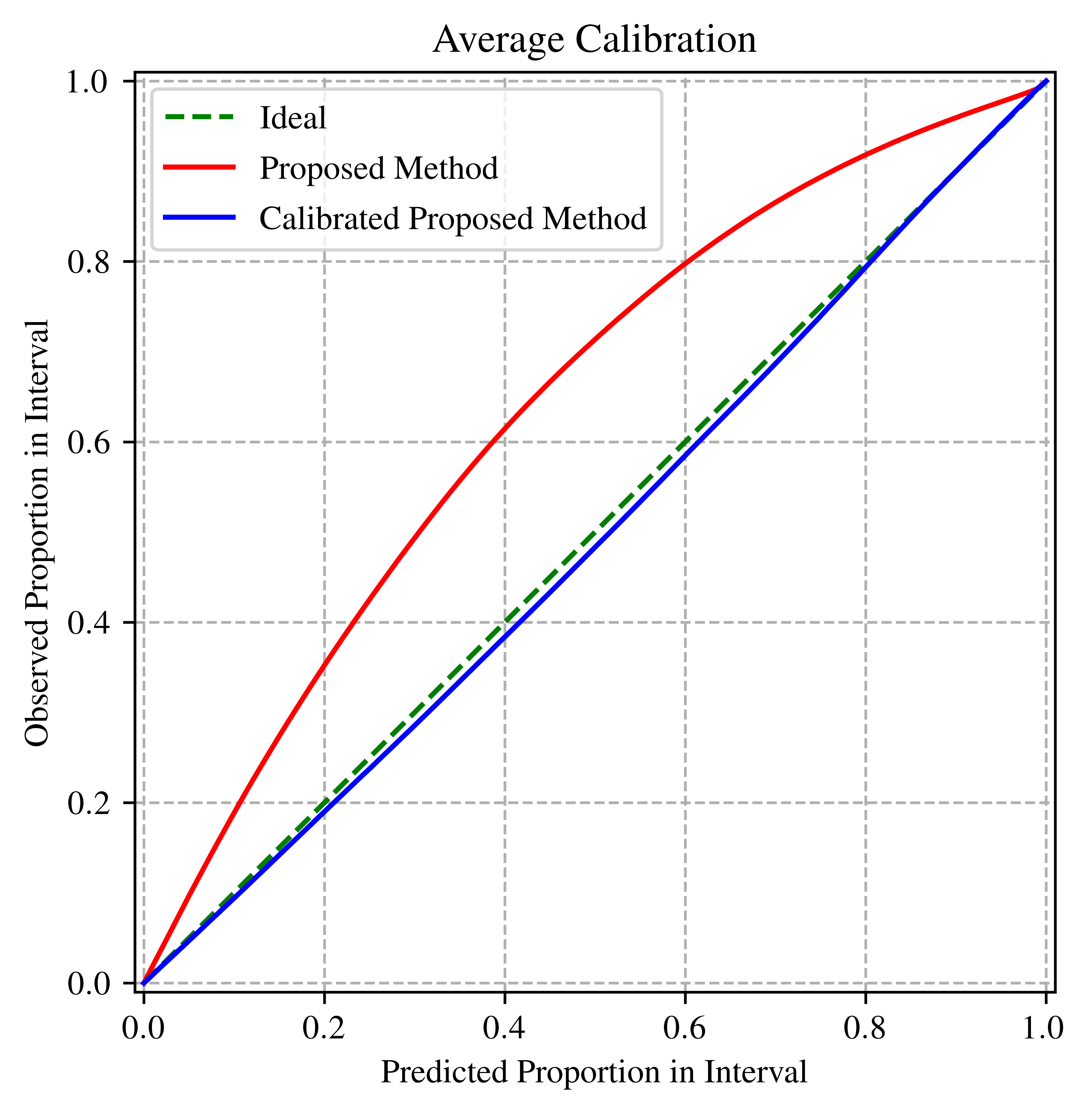}
    }
    \hfill
   \subfloat[$36$, $70$]{%
      \includegraphics[width=0.45\columnwidth]{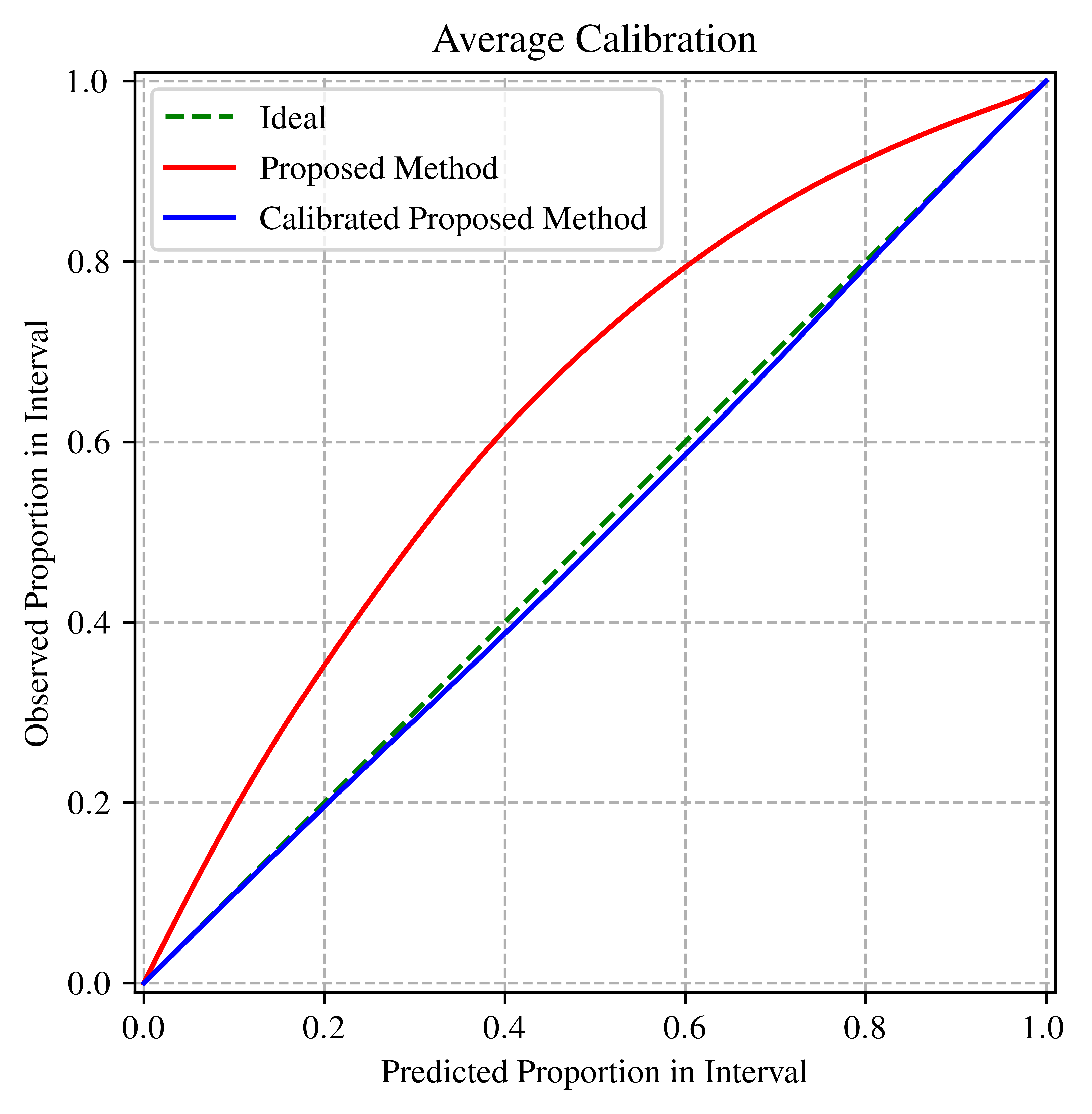}
    } \\
    \subfloat[$60$, $40$]{%
      \includegraphics[width=0.45\columnwidth]{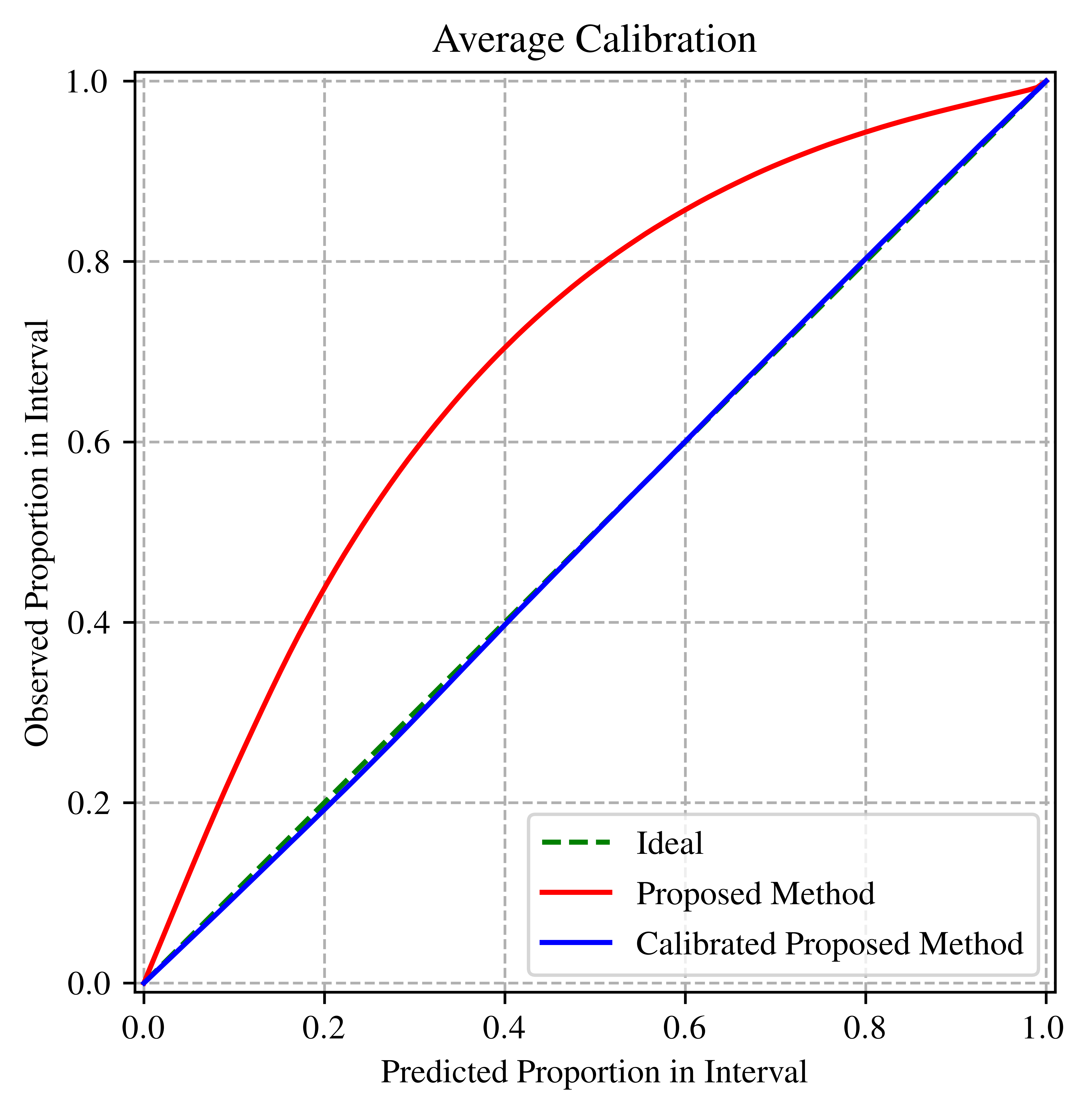}
    }
    \hfill
    \subfloat[$60$, $70$]{%
      \includegraphics[width=0.45\columnwidth]{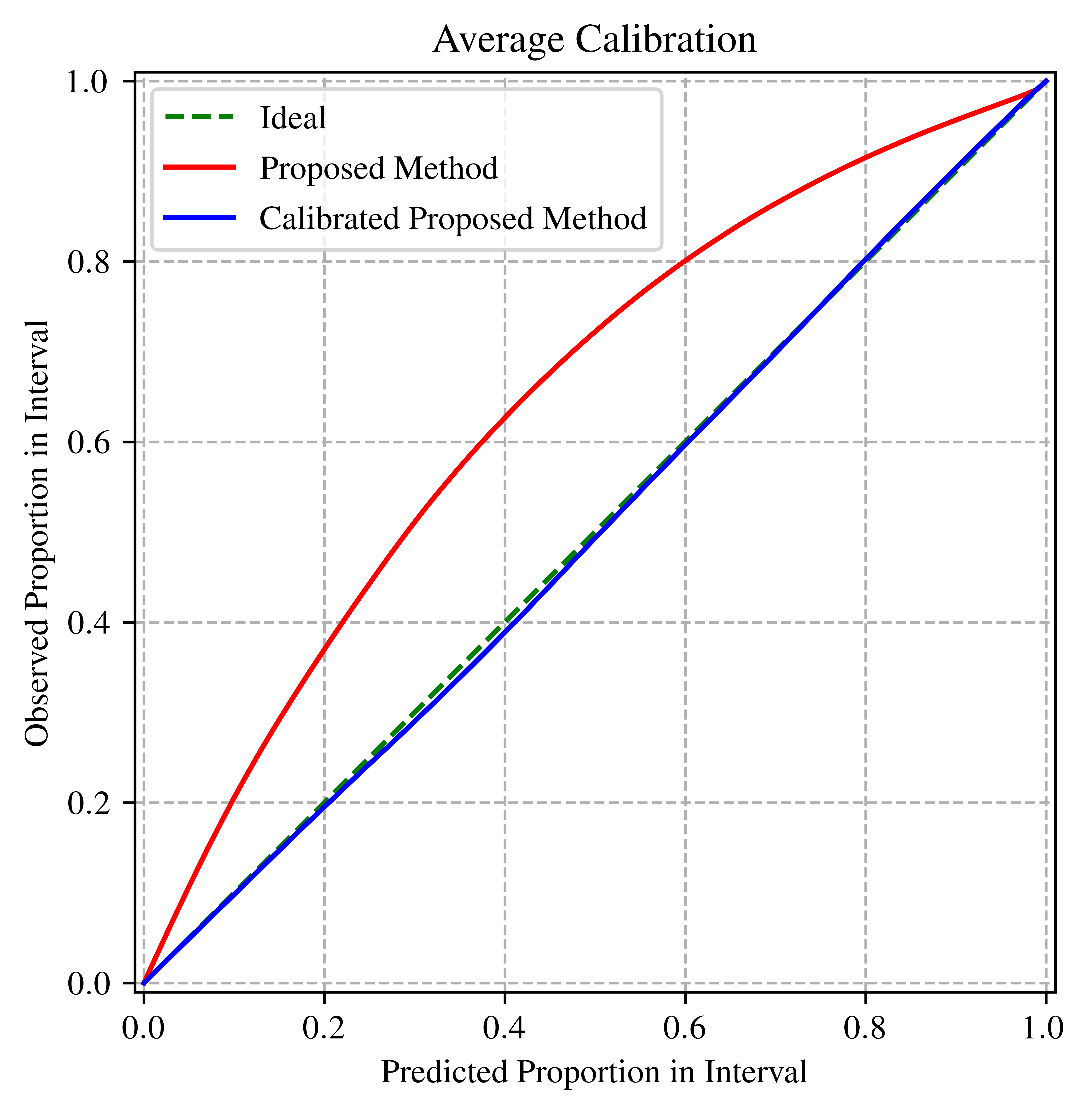}
    }
    \caption{Calibration plots of the proposed method for different configurations of the CT setup. Subcaptions specify the number of views and the SNR (dB).}
    \label{fig:calibration_plots_ct}
\end{figure}

A calibration plot is a diagnostic tool that allows visually inspecting the calibration properties of a probabilistic model to understand whether the model is providing underconfident or overconfident predictions. In this section, to obtain calibration plots, we approximate the predictive distribution of the proposed method, which is a mixture of Gaussians distribution with $T$ mixture components, with a multivariate Gaussian distribution. More specifically, we approximate the predictive distribution of each pixel with a Gaussian distribution as follows:
\begin{equation}
p([\s_*]_k | \m_*, \mathcal{D}) \approx  \mathcal{N}([\s_*]_k| [\mathbb{E}[\s_* | \m_*, \mathcal{D}]]_k, \Var[[\s_*]_k | \m_*, \mathcal{D}]),
\label{eq:preddistrecalibration}
\end{equation}
where the mean and the variance of the distribution are defined in \eqref{eq:predictivemean} and \eqref{eq:predictivevariance}, respectively. Using this approximation, we generated calibration plots on the test datasets using Uncertainty Toolbox~\cite{chung2021uncertainty} for different configurations of the MRI and CT observation models. Figure \ref{fig:calibration_plots_mri} and Figure \ref{fig:calibration_plots_ct} show those calibration plots for the MRI and CT experiments, respectively. The red curves in the plots represent the calibration curves of the proposed method for different configurations of the imaging setups. The dashed green line, on the other hand, represents the \emph{ideal} calibration curve. 

Figure \ref{fig:calibration_plots_mri} and Figure \ref{fig:calibration_plots_ct} show that the proposed model may provide slightly underconfident predictions. 
The main reason behind this bias, which is sometimes referred to as the model bias, is the assumptions we have made about the form of the likelihood function, the prior distribution of the parameters of the likelihood function, and the choice of the parametric distribution that we use to approximate the true posterior distribution of the parameters. Luckily, we can easily \emph{recalibrate} the proposed method by following the recalibration method introduced by Kuleshov \emph{et~al.}~\cite{Kulesov2018recalibration}. In our experiments, we used the validation dataset as the calibration dataset to recalibrate the proposed method. Calibration curves of the calibrated proposed method are represented with the blue color in Figure \ref{fig:calibration_plots_mri} and Figure \ref{fig:calibration_plots_ct}. After recalibration, we observe that the calibrated proposed model is capable of outputting more calibrated probabilistic predictions.

\subsection{Reconstruction Performance}
\label{ssec:reconstructionperformance}
In this subsection, we demonstrate the reconstruction performance of the proposed framework. We compare the proposed framework with six methods: (1) zero-filling (ZF) / filtered backprojection (FBP), (2) total variation reconstruction (TV), (3) PGD-based deep unrolling method (PUM), (4) PGD-based deep unrolling method without batch normalization (PUMw/oBN), (5) proposed only epistemic model (POEM), and (6) proposed only aleatoric model (POAM).

The methods ZF/FBP, and TV are the baseline reconstruction methods that we use to demonstrate how challenging the reconstruction problem is. PUM is a deep unrolling method using PGD. Each residual block of PUM consists of a series of convolutional layers, batch normalization layers, and an activation function. PUMw/oBN is the same model as the PUM, except that there are no batch normalization layers in residual blocks. POEM is the variant of the proposed framework that assumes that the covariance matrix of the likelihood function in \eqref{eq:proposedlikelihood} is a fixed model parameter. POEM is also the probabilistic model that was used in the experiments of the preliminary version of this paper~\cite{Ekmekci2021UncertaintyUnfoldingPreliminary}. As its name implies, POEM quantifies only the epistemic uncertainty, not the aleatoric uncertainty. POAM is also a variant of the proposed framework where a maximum likelihood estimate of the parameters of the likelihood function in \eqref{eq:proposedlikelihood} is used. POAM is capable of quantifying the aleatoric uncertainty, but not the epistemic uncertainty since it only uses the MAP estimate of the parameters. Implementation details of these methods are provided in the Supplementary Material.

Table \ref{tab:reconstructionperformance} shows the performance of the seven methods for CT and MRI reconstruction problems under different configurations of the imaging setups. Among these image reconstruction methods, FBP and ZF achieve the worst reconstruction performance among the seven reconstruction methods. The TV method improves upon FBP and ZF by promoting a piecewise-constant reconstruction. The deep unrolling method PUM surpasses the TV method by implicitly learning the prior using the training dataset. The deep unrolling method PUM was trained using a small mini-batch size since it requires storing the intermediate variables having the same spatial dimensions as the target image in the memory to carry out the backpropagation. We empirically observed that the removal of the batch normalization layers from the unrolled network leads to an increase in the reconstruction performance. Specifically, we observe that the PUMw/oBN outperforms PUM in all the experiments. This empirical observation is mathematically justified in \cite{Yong2020BatchNormalization} where Yong \emph{et. al.}\ showed that batch normalization introduces a high level of noise for small mini-batch sizes, making the training difficult. This observation is the main reason why the unrolled network $f$ in the proposed framework does not contain any batch normalization layers. On the other hand, we experimentally observed that the addition of the batch normalization layers into the neural network $\sigma^2$ is necessary to have a stable training stage. Comparing POAM with PUMw/oBN, POAM shows an average SSIM decrease of $0.022$ for the MRI reconstruction problem and $0.002$ for the CT reconstruction problem. On the other hand, when compared to the state-of-the-art deep unrolling method PUM, POAM achieves average SSIM gains of $0.031$ and $0.011$ for the MRI and CT reconstruction problems, respectively. The reconstruction performance of POEM shows a decrease compared to PUMw/oBN due to using dropout after every convolutional layer, which is a strong form of regularization. Similarly, we observe that the reconstruction performance of POEM is slightly worse than that of POAM. The reconstruction performance of the proposed framework shows a decrease compared to POAM because of using dropout after every convolutional layer, which is a strong form of regularization. Comparing the proposed framework with POAM, the proposed framework shows an average SSIM decrease of $0.010$ for the MRI reconstruction problem and $0.007$ for the CT reconstruction problem. We observe a similar trend for the proposed framework and PUMw/oBN. On the other hand, the proposed framework achieves average SSIM gains of $0.010$ and $0.006$ for the MRI and CT reconstruction problems when compared to POEM, respectively. Similarly, the proposed framework surpasses the state-of-the-art deep unrolling method PUM. Due to space limitations, only representative visual results are presented in Figure \ref{fig:reconstructionperformancevisuals}. Detailed visual results are provided in the Supplementary Material.

\begin{figure*}[t]
    \centering
    \includegraphics[width=0.7\textwidth]{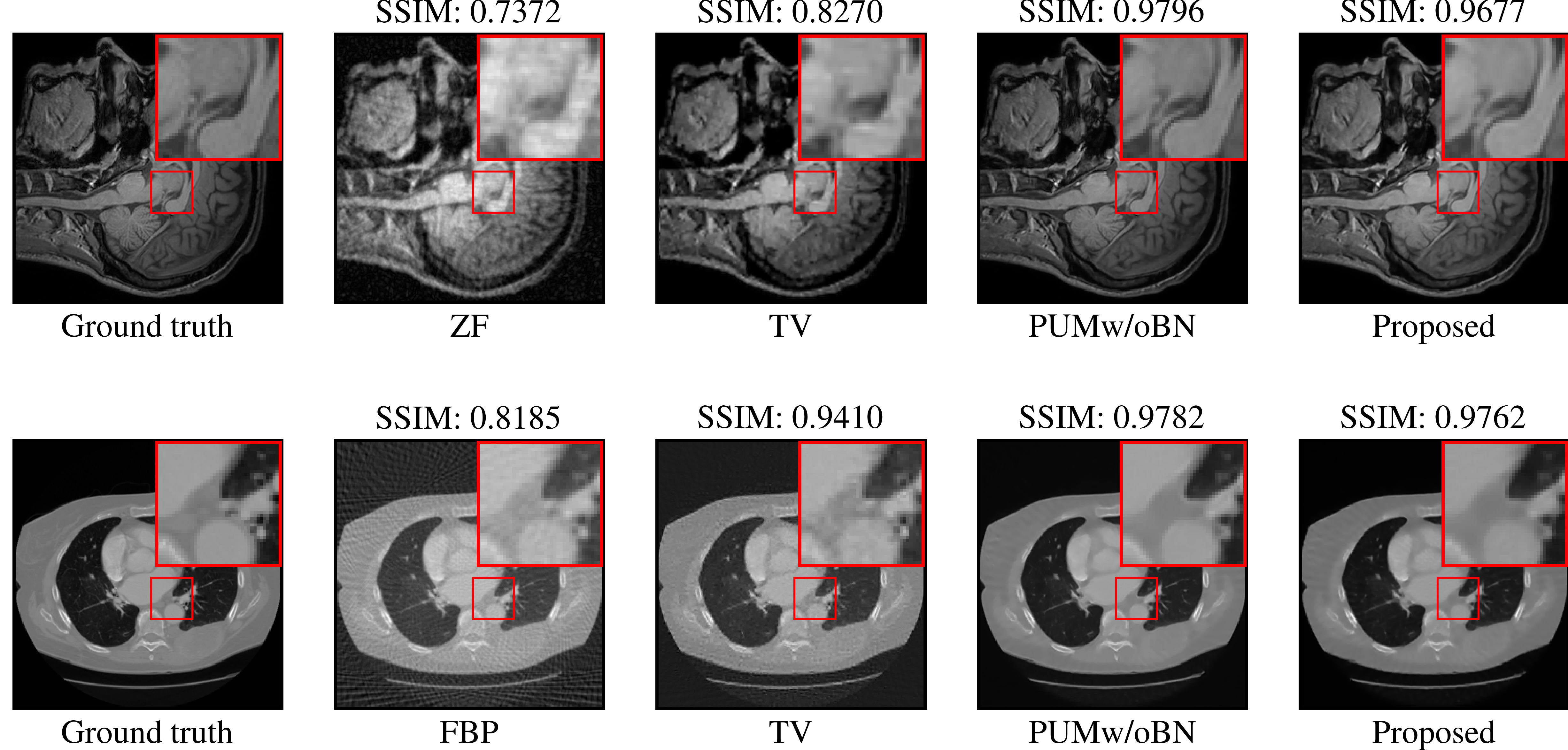}
    \caption{Visual comparison of the image reconstruction performance of zero-filling (ZF) / filtered backprojection (FBP), total variation reconstruction (TV), state-of-the-art PGD-based deep unrolling method without batch normalization (PUMw/oBN), and the proposed method. Proposed method achieves comparable reconstruction performance to the state-of-the-art deep unrolling method PUMw/oBN, while providing uncertainty quantification.}
    \label{fig:reconstructionperformancevisuals}
\end{figure*}

\begin{table*}[t]
\centering
\caption{Comparison of average SSIM for different image reconstruction methods.}
\label{tab:reconstructionperformance}
\begin{tabular}{cccc|cccccccc}
\toprule
& POC & NOV & SNR & ZF & FBP & TV & PUM & PUMw/oBN & POAM & POEM & Proposed \\
\midrule
\multirow{4}{*}{MRI} & 10 & - & 10 & 0.4910 & - & 0.7033 & 0.7979 & 0.8144 & 0.7773 & 0.7727 & 0.7674 \\
& 10 & - & 70 & 0.5448 & - & 0.7261 & 0.8032 & 0.9227 & 0.8996 & 0.8638 & 0.8784 \\
& 20 & - & 10 & 0.5609 & - & 0.7913 & 0.8611 & 0.8799 & 0.8589 & 0.8517 & 0.8568 \\
& 20 & - & 70 & 0.6774 & - & 0.8414 & 0.9231 & 0.9780 & 0.9726 & 0.9407 & 0.9642 \\
\midrule
\multirow{4}{*}{CT} & - & 36 & 40 & - & 0.4919 & 0.7657 & 0.9053 & 0.9228 & 0.9178 & 0.9068 & 0.9129 \\
& - & 36 & 70 & - & 0.5895 & 0.8232 & 0.9175 & 0.9319 & 0.9290 & 0.9133 & 0.9181 \\
& - & 60 & 40 & - & 0.6726 & 0.8637 & 0.9390 & 0.9535 & 0.9520 & 0.9422 & 0.9467 \\
& - & 60 & 70 & - & 0.7846 & 0.9204 & 0.9548 & 0.9625 & 0.9626 & 0.9507 & 0.9576 \\
\bottomrule
\end{tabular}
\end{table*}

\section{Discussion}
\label{sec:discussion}
Quantification of the epistemic uncertainty is crucial for learning-based image reconstruction methods, especially in safety-critical imaging applications, for quantifying the confidence on a reconstruction obtained using a model learned from available, potentially limited or unrepresentative training data. Our experimental results presented in Section \ref{sec:experiments} showed that the epistemic uncertainty information provided by the proposed method exhibits the reducibility property of the epistemic uncertainty. Moreover, the epistemic uncertainty provided by the proposed framework can be used to assess how uncertain the learning-based image reconstruction method is and to detect cases where the input contains abnormal features not present in the training data.

For ill-posed inverse problems encountered in most imaging problems, inherent uncertainty on the target image for a given measurement vector is inevitable. Hence, it is essential to quantify the aleatoric uncertainty for imaging problems to capture the inherent randomness in the reconstruction task. Our experiments presented in Section \ref{sec:experiments} demonstrated that the proposed framework is capable of capturing the aleatoric uncertainty in the sense that the aleatoric uncertainty provided by the proposed framework shows the regions where the reconstruction is expected to be challenging for the unrolled
network. The aleatoric uncertainty provided by the proposed framework can be utilized to determine the possible errors in the reconstructed image and can be used as a mechanism to further assess the reliability of the reconstructed image. As a result, the aleatoric and epistemic uncertainties provided by the proposed framework would open the possibility of developing more accurate, robust, trustworthy, uncertainty-aware, learning-based image reconstruction and analysis methods. While the uncertainty estimates provided by our proposed methodology appear to reflect expected behavior of epistemic and aleatoric uncertainties, further analysis of the implications of the variational inference approximations used here would be beneficial.

The benefits of obtaining the epistemic and aleatoric uncertainty maps come with a price. Because the proposed framework requires feeding the measurement vector into the neural networks $T$ times for inference, the inference time of the proposed framework increases by $T$ times compared to the state-of-the-art deep unrolling method PUM. To shorten the inference time of the proposed framework, we can perform those $T$ forward passes in parallel. Assuming that the GPU memory allows using a batch size of $B$ in the inference stage, the proposed framework requires only $\lfloor T/B \rfloor + 1$ forward passes for inference. If we have multiple GPUs, the inference time of the proposed framework can be further reduced. Hence, the proposed framework can achieve shorter inference times at the expense of using more computational power. Another way to shorten the inference time of the proposed framework is to decrease the number of parameters the proposed framework so that a larger batch size $B$ can be used to parallelize the inference stage. To that end, we can design a variant of the proposed framework that uses a dual-head network. For the sake of brevity, we have not discussed this variant; however, a brief discussion on that variant is provided in the Supplementary Material.

\section{Conclusion}
\label{sec:conclusion}
In this paper, we utilized the idea of deep unrolling and Bayesian neural networks to propose a learning-based image reconstruction framework that is capable of quantifying epistemic and aleatoric uncertainties while incorporating the imaging observation model into the reconstruction process. Our experimental results showed that the proposed framework provides epistemic and aleatoric uncertainty maps while providing a reconstruction performance comparable to the state-of-the-art deep unrolling methods. The proposed framework can be applied to a broad set of imaging problems and can be easily implemented in deep learning frameworks. We hope that the proposed framework and the provided discussion on epistemic and aleatoric uncertainties for imaging problems motivate further research on uncertainty characterization for imaging problems and on leveraging the uncertainty information for image reconstruction and analysis tasks.

\bibliographystyle{IEEEtran}
\bibliography{refs_main}

\vfill

\end{document}


\title{Supplementary Material: Uncertainty Quantification for Deep Unrolling-Based Computational Imaging}

\author{Canberk Ekmekci,~\IEEEmembership{Student Member,~IEEE}, and Mujdat Cetin,~\IEEEmembership{Fellow,~IEEE}
}

\markboth{Journal of \LaTeX\ Class Files,~Vol.~14, No.~8, August~2021}%
{Shell \MakeLowercase{\textit{et al.}}: A Sample Article Using IEEEtran.cls for IEEE Journals}


\maketitle

\section{Details of the Neural Networks $f$ and $\sigma^2$}
\label{sec:detailsofournetwork}
\begin{figure}[t]
	\centering
	\includegraphics[width=\columnwidth]{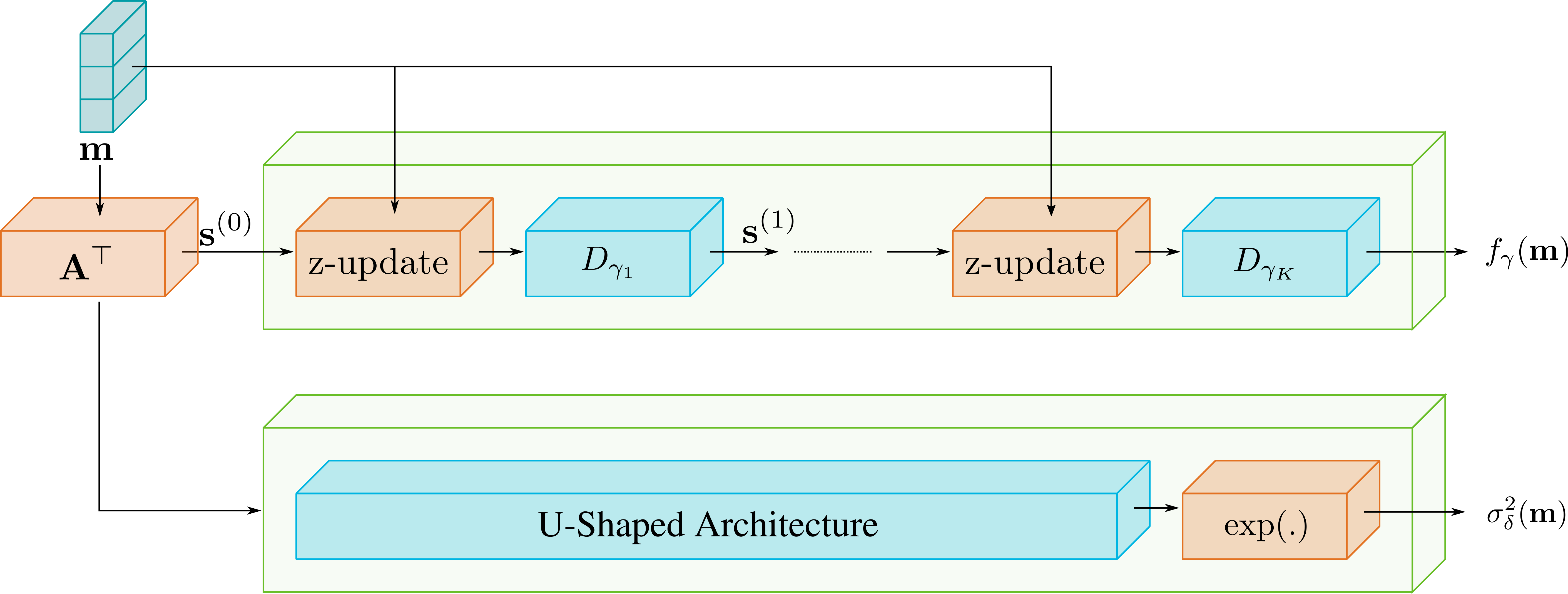}
	\caption{High level overview of the neural networks that define the form of the Gaussian likelihood function of the proposed framework.}
	\label{fig:highlevel}
\end{figure}

\begin{figure}[t]
	\centering
	\includegraphics[width=\columnwidth]{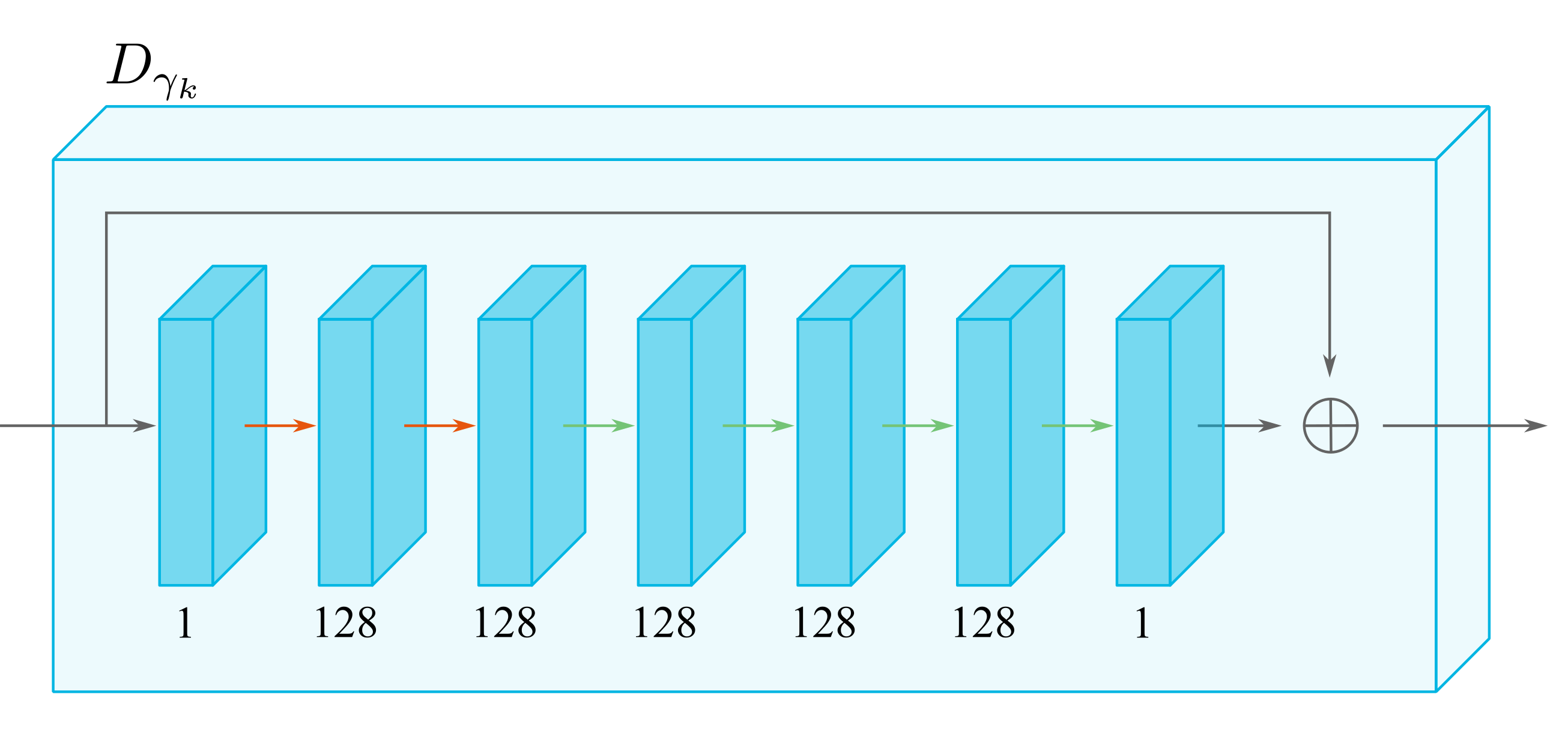}
	\caption{Details of a residual block $D_{\gamma_k}$ used in the neural network $f$ for CT reconstruction. For MRI reconstruction, the number of input and output channels is two instead of one. Red arrows represent convolutional layers with a kernel size of $3\times 3$ and a padding size of $1$ followed by a LeakyReLU activation function. Green arrows represent convolutional layers with a kernel size of $1\times1$ and a padding size of $0$ followed by a LeakyReLU activation function.}
	\label{fig:residual}
\end{figure}

\begin{figure}[t]
	\centering
	\includegraphics[width=\columnwidth]{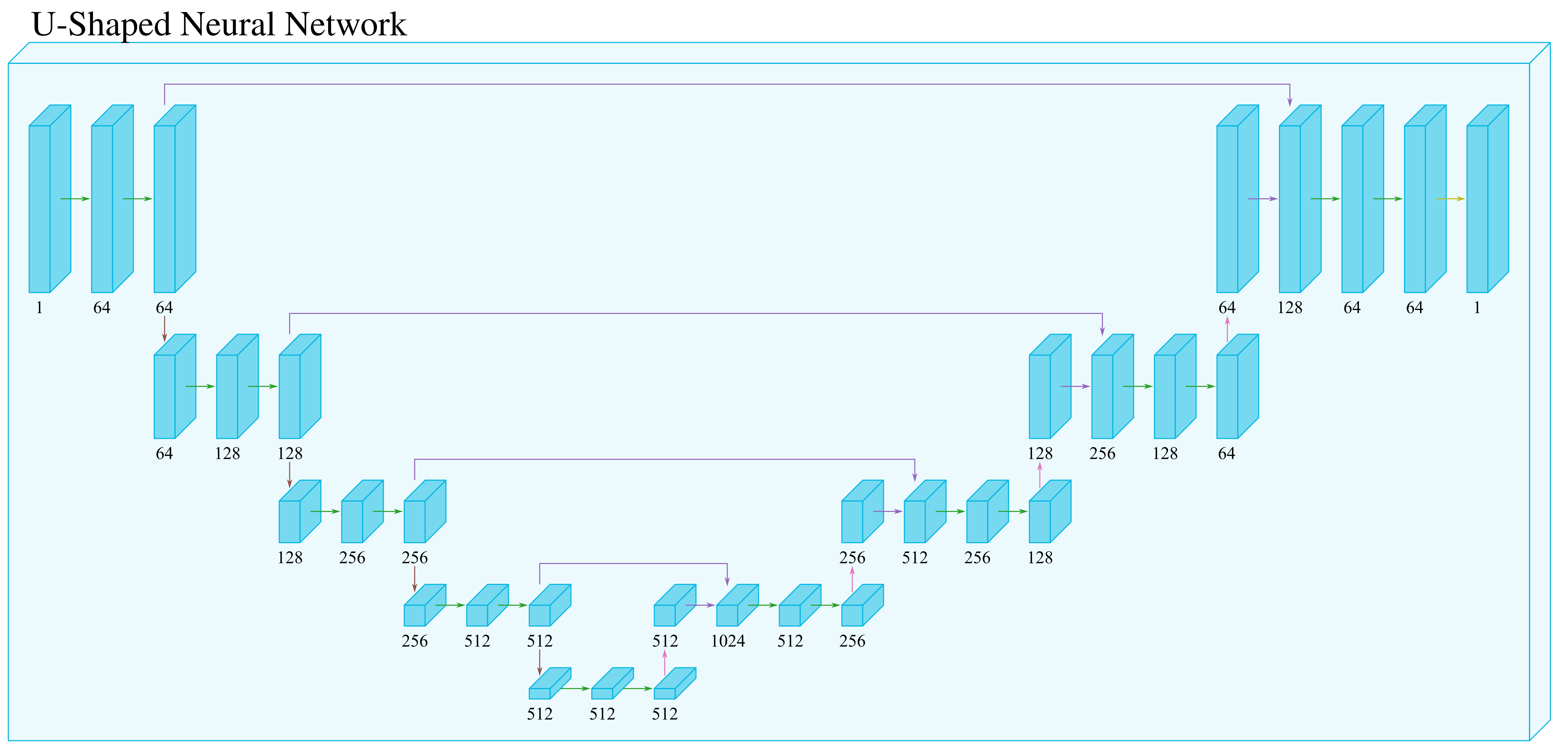}
	\caption{Details of the U-Shaped neural network used in the neural network $\sigma^2$ for CT reconstruction. For MRI reconstruction, the number of input and output channels is two instead of one. Green arrows represent convolutional layers with a kernel size of $3 \times 3$ and a padding size of $1$ followed by batch normalization and ReLU activation function. Brown arrows represent maxpooling layers with a kernel size of $2$ and a stride of $2$. Pink arrows represent bilinear upsampling with a scale factor of $2$. Purple arrows represent the concatenation operation along the channel dimension. Yellow arrows represent convolutional layers with a kernel size of $3\times 3$ and a padding size of $1$.}
	\label{fig:unet}
\end{figure}

Figure \ref{fig:highlevel} shows the neural network architectures that define the Gaussian likelihood function of the proposed framework at a high level. Architectural details of the residual blocks~\cite{He2016Resnet} used in the deep unrolled network $f$ are illustrated in Figure \ref{fig:residual}. Mardani \emph{et~al.}\ \cite{Mardani2018NPGD} have used the same architecture for the residual blocks to develop a proximal gradient descent-based deep unrolling method for MRI reconstruction. Architectural details of the neural network $\sigma^2$, which models the diagonal entries of the covariance matrix of the Gaussian likelihood function, are shown in Figure \ref{fig:unet}.

We emphasize that the proposed framework uses MC Dropout~\cite{Srivastava2014Dropout} to approximate the posterior distribution of the parameters of the likelihood function, which requires the use of dropout~\cite{Srivastava2014Dropout} after the convolutional layers of the neural networks $f$ and $\sigma^2$. We use dropout after every convolutional layer, except the first convolutional layer, of every residual block to obtain the dropout-added neural network $\bar{f}$. Similarly, to obtain the dropout-added neural network $\bar{\sigma}^2$, we use dropout after every convolutional layer of the neural network $\sigma^2$. 

\begin{figure}[t]
	\centering
	\includegraphics[width=\columnwidth]{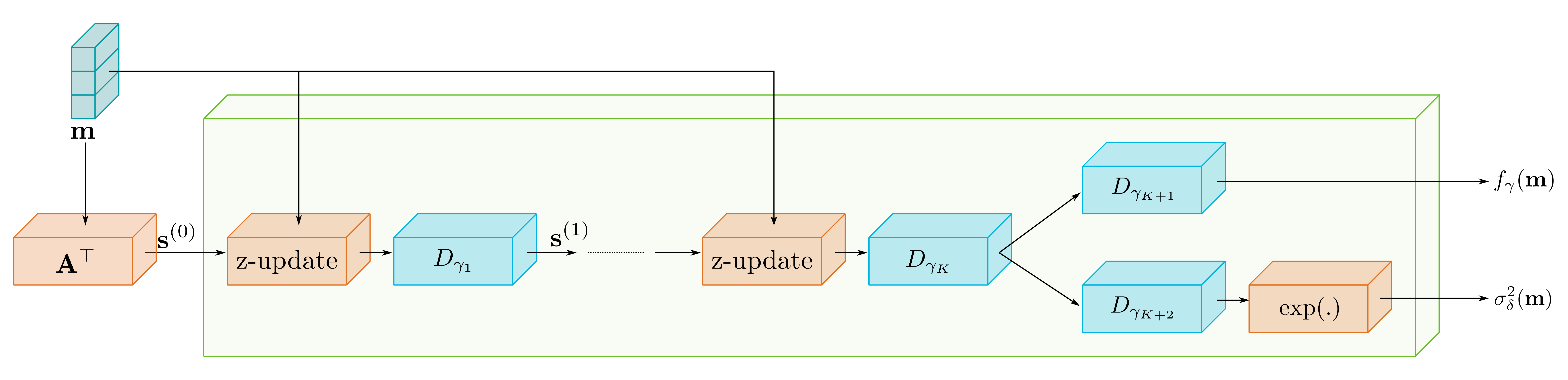}
	\caption{High level overview of the dual-head neural network that simultaneously outputs the mean and the covariance matrix of the Gaussian likelihood function of the proposed method.}
	\label{fig:highleveldualhead}
\end{figure}

\begin{figure}[t]
	\centering
	\includegraphics[width=\columnwidth]{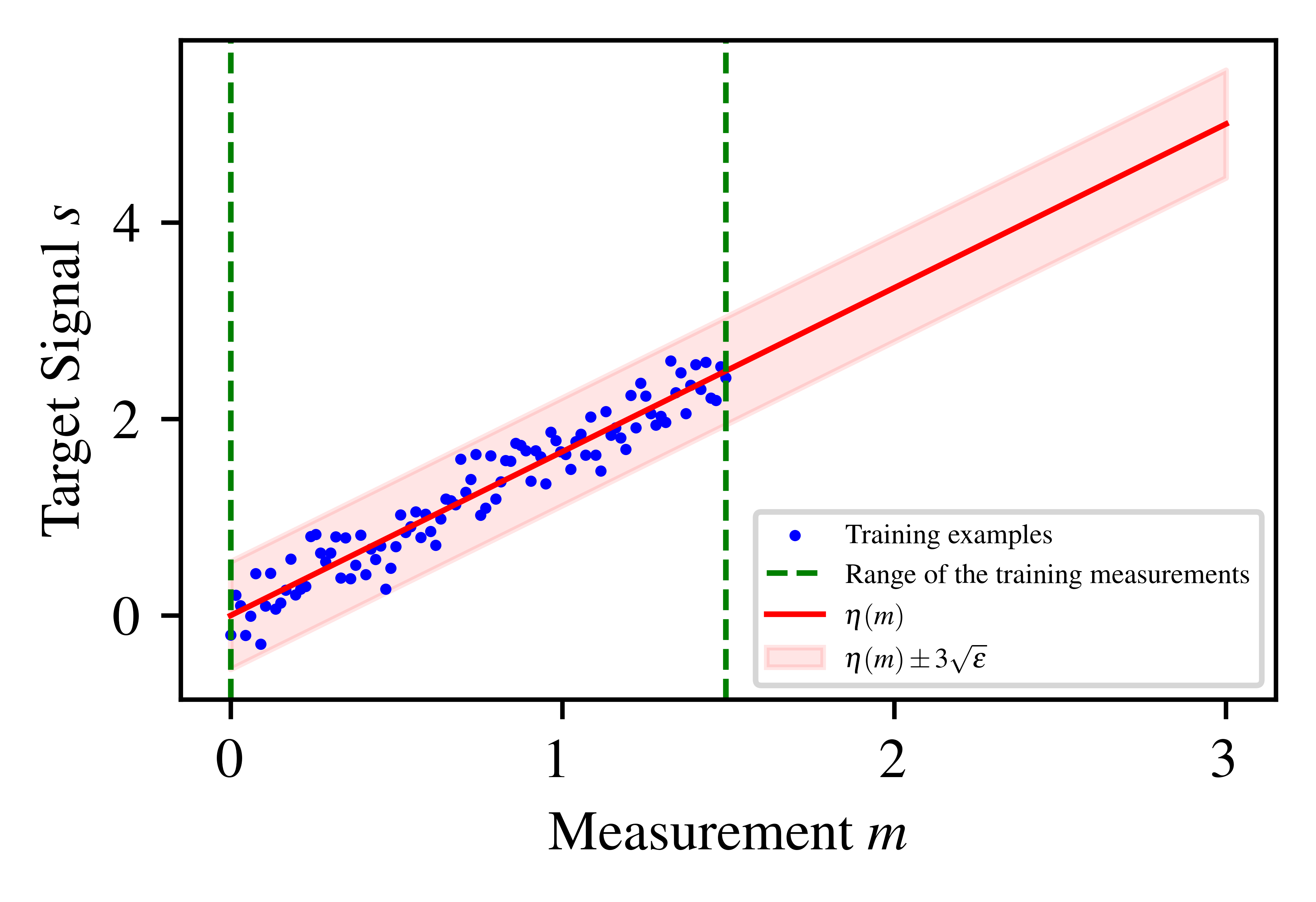}
	\caption{Details of the one dimensional toy problem.}
	\label{fig:ps}
\end{figure}

\section{Dual-Head Architecture}
\label{sec:dualhead}
In the proposed framework, we have used a U-shaped architecture~\cite{Ronneberger2015UNet} to represent the covariance matrix of the Gaussian likelihood function of the proposed method. An alternative is to use a dual-head architecture~\cite{Kendall2017BayesianNN}, which is illustrated in Figure \ref{fig:highleveldualhead}. The advantage of using a dual-head architecture is that it is less GPU memory intensive, and larger batch sizes can be used for the training and inference stages, leading to faster inference time. For example, the proposed framework presented in the paper allows the use of a batch size of $4$ on a 16GB GPU, whereas the dual-head variant of the proposed framework allows the use of a batch size of $5$, leading to a decrease in the average inference time from $5.30$ seconds to $4.65$ seconds. On the other hand, we have experimentally observed that training a dual-head architecture is slightly more challenging compared to the proposed method presented in the paper and that aleatoric uncertainty maps obtained by the dual-head variant of the proposed method are noisier compared to ones obtained by the proposed framework presented in the paper.

\section{Toy Problem}
\label{sec:toyproblem}
In this section, we provide a one dimensional inverse problem, which we refer to as the toy problem, to make the abstract concepts of aleatoric and epistemic uncertainties more concrete and to show that the proposed framework successfully captures epistemic and aleatoric uncertainties.

As a toy problem, we consider a one dimensional linear inverse problem having the form 
\begin{equation}
m = a s + n,
\end{equation}
where $m \in \mathbb{R}$ is the measurement, $a \in \mathbb{R}$ is the forward operator, $s \in \mathbb{R}$ is the target signal, and $n \sim \mathcal{N}(0, \sigma_n^2)$ is additive white Gaussian noise. For this setup, we choose the true prior distribution of the target signal $s$ to be $p(s) = \mathcal{N}(s| \mu, \tau^{-1})$. Thus, the posterior distribution of the target signal given a measurement $m$ becomes
\begin{equation}
    p(s|m) = \mathcal{N}(s | \eta(m), \epsilon),
\label{eq:posterior}
\end{equation}
where $\eta(m)=\epsilon[a \sigma_{n}^{-2}m + \tau \mu]$, and $\epsilon = (\tau + a^2 \sigma_{n}^{-2})^{-1}$. 

For the experiment, we chose the following values for the parameters of the toy problem: $a=0.5, \sigma_n=0.1, \mu=0$, and $ \tau^{-1}=0.2$. We obtained the training dataset by taking $100$ measurements uniformly spaced over the interval $[0,3/2]$ and generating the corresponding target signals by sampling from the distribution $p(s|m)$. Figure \ref{fig:ps} shows the details of the toy problem. 

\begin{figure}[t]
	\centering
	\includegraphics[width=\columnwidth]{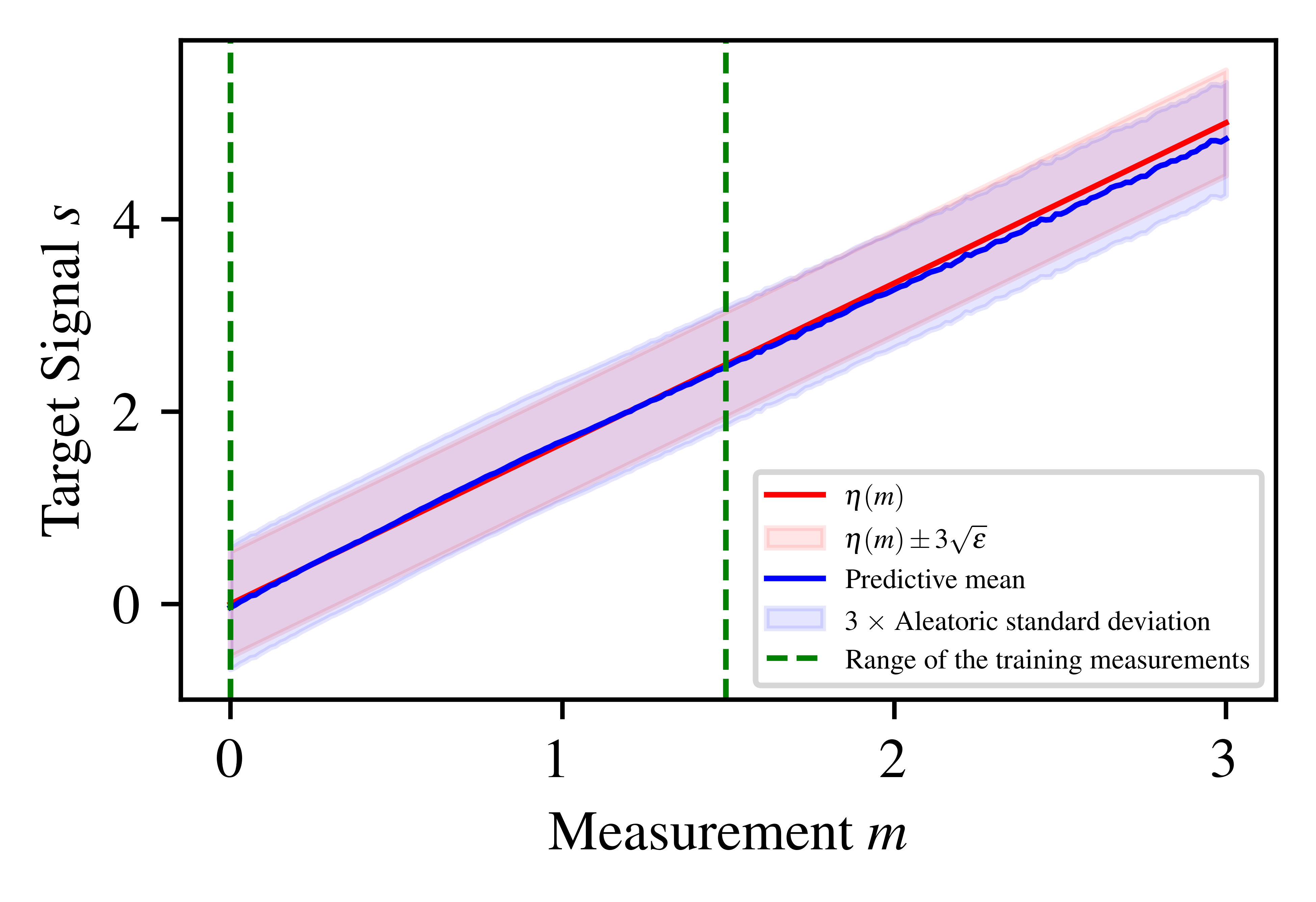}
	\caption{True aleatoric uncertainty and the aleatoric uncertainty captured by the proposed framework for the toy problem.}
	\label{fig:aleatoric}
\end{figure}

\begin{figure}[t]
	\centering
	\includegraphics[width=\columnwidth]{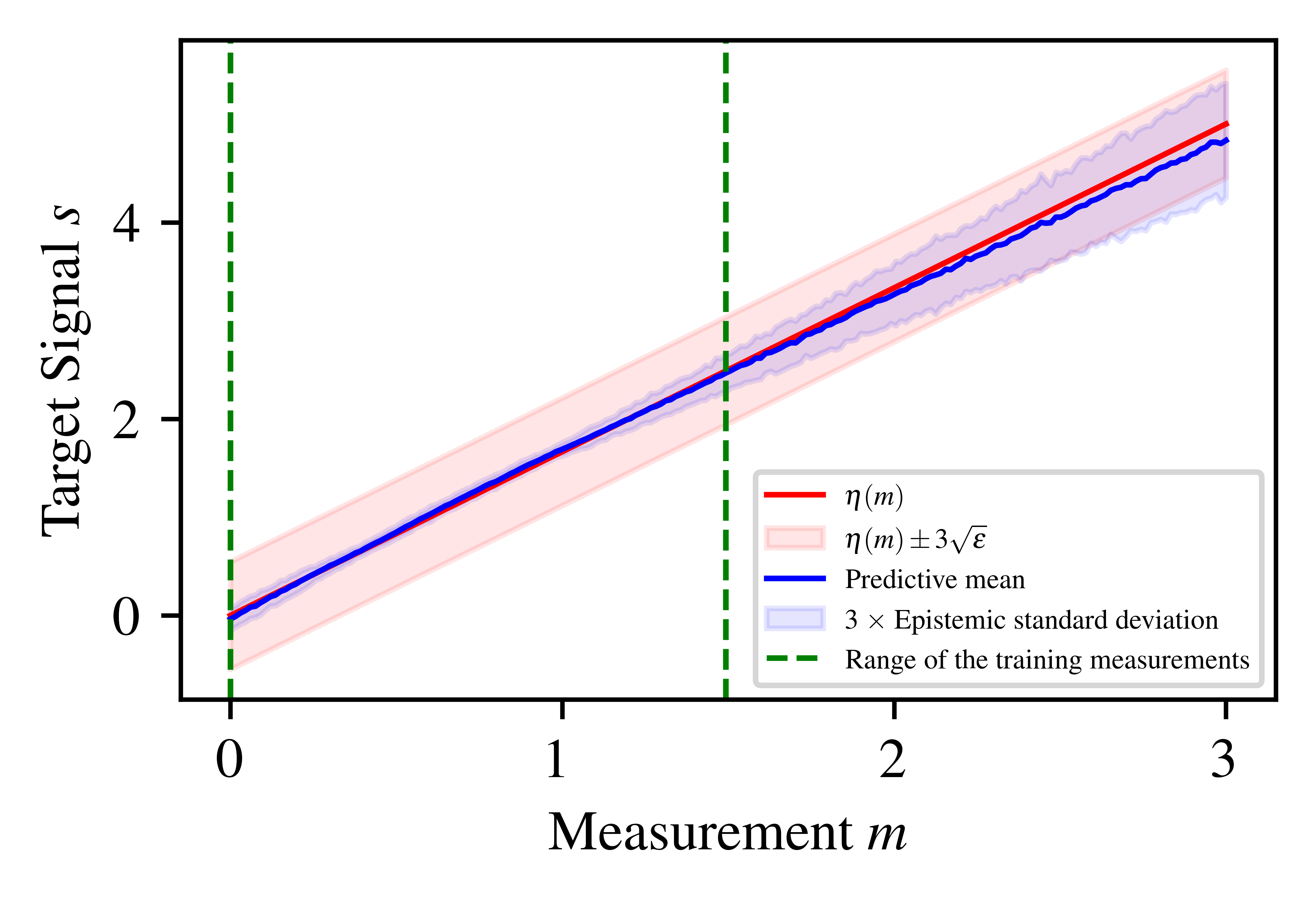}
	\caption{Epistemic uncertainty captured by the proposed framework.}
	\label{fig:epistemic}
\end{figure}

For this toy problem, we used the proposed framework to obtain epistemic and aleatoric uncertainties. We used multi-layer perceptrons (MLPs) with three hidden layers for the residual blocks of the neural network $f$ and a MLP with two hidden layers for the neural network $\sigma^2$. Because the output is one dimensional, we did not put a dropout layer at the end of the neural networks. Initial step size of the neural network $f$ was fixed to $1.0$, and number of iterations $K$ of the proximal gradient descent was set to $5$. The dropout rate was fixed to $0.5$, and the proposed framework was trained for $20000$ epochs using the learning rate of $1 \times 10^{-4}$. In the inference stage, for $200$ uniformly spaced test measurement vectors over the interval $[0,3]$, we computed the reconstruction, aleatoric standard deviation and the epistemic standard deviation. Figure \ref{fig:aleatoric} and Figure \ref{fig:epistemic} show the aleatoric and epistemic uncertainties captured by the proposed framework, respectively.

For this toy problem, the inherent uncertainty in the reconstruction task, i.e., the uncertainty on the target signal for a given measurement, is caused by the variance term $\epsilon$ (see \eqref{eq:posterior}). For the test measurements that lie in the interval $[0,3/2]$, which matches the interval of the training dataset, the aleatoric uncertainty captured by the proposed framework significantly overlaps with the true aleatoric uncertainty. Epistemic uncertainty on the other hand is the uncertainty on the parameters, which is due to lack of training examples around a test measurement. For the test measurements that lie in the interval $[0,3/2]$, which matches the interval of the training dataset, the epistemic uncertainty captured by the proposed framework is low as we expected. As we move towards to the region for which we do not have any training data, i.e., as the measurements start deviating from the training data, the epistemic uncertainty captured by the proposed framework increases.

\section{Other Methods Used in Reconstruction Experiments}
\label{sec:othermethods}
In the experiments section of the article, we have compared the reconstruction performance of the proposed framework with six other reconstruction methods. Descriptions of those methods and their implementation details are provided below.

\textbf{Zero-Filling:} Zero-filling is one of the baseline reconstruction methods that we have used for the MRI experiments. This method simply fills the unobserved Fourier (k-space) coefficients with zeros and computes the inverse Fourier transform.

\textbf{Filtered Backprojection:} Filtered backprojection is one of the baseline reconstruction methods that we have used for the CT experiments. This method first filters the sinogram data and then computes the backprojection of the filtered sinogram. In our experiments, we used the TorchRadon~\cite{torch_radon} package to implement this method.

\textbf{Total Variation:} Total variation reconstruction is the second baseline reconstruction method that we have used in our experiments. This method solves the following optimization problem to reconstruct the image.
\begin{equation}
            \hat{\s} = \argmin_{\s} \left\{ \| \A \s - \m \|_2^2 + \beta \| \s \|_{\text{TV}} \right\},
\end{equation}
where $\|.\|_{\text{TV}}$ denotes the total variation semi-norm~\cite{Chambolle2004TV}. We have used alternating direction method of multipliers (ADMM)~\cite{Boyd2011ADMM} to obtain an iterative algorithm that solves this optimization problem. In our experiments, the number of iterations and the penalty parameter of the ADMM were fixed to $100$ and $10.0$, respectively. The data-dependent update step of the ADMM was solved using conjugate gradient (CG) method. The tolerance parameter of the CG was fixed to $1 \times 10^{-5}$, and maximum number of CG iterations was set to $10$. The value of the regularization constant $\beta$ was chosen from the set $\{1 \times 10^{-4}, 1 \times 10^{-3}, 1 \times 10^{-2}, 1 \times 10^{-1}, 1 \times 10^{0}, 1 \times 10^{1} \}$ to maximize the SSIM.

\textbf{Deep Unrolling:} This is a learning-based image reconstruction method that leverages the idea of deep unrolling. The neural network used for this method is the neural network $f$ depicted in Figure \ref{fig:highlevel} with the residual blocks in Figure \ref{fig:residual}, except that there is a batch normalization layer between every convolutional layer and the activation function ReLU. The batch size was set to $4$ for MRI experiments and $2$ for the CT experiments, and learning rate was set to $1\times 10^{-4}$ for the MRI experiments and $1\times 10^{-5}$ for the CT experiments. Initial step size of the neural network $f$ was set to $1.0$ for the MRI experiments and $1\times 10^{-4}$ for the CT experiments. The number of iterations of the PGD was fixed to $5$, and the neural network was trained for $100$ epochs using mean-squared error loss function.

\textbf{Deep Unrolling without Batch Normalization:} This is another variant of the deep unrolling method used in the reconstruction experiments. The neural network used for this method is the neural network $f$ depicted in Figure \ref{fig:highlevel} with the residual blocks in Figure \ref{fig:residual}. The batch size was set to $4$ for the MRI experiments and $2$ for the CT experiments. Learning rate was fixed to $1\times 10^{-4}$ for the MRI experiments and $1\times 10^{-5}$ for the CT experiments. Initial step size of the neural network $f$ was set to $1.0$ for the MRI experiments and $1\times 10^{-4}$ for the CT experiments. The number of iterations of the PGD was set to $5$, and the neural network was trained for $100$ epochs using mean-squared error loss function.

\textbf{Proposed Only Aleatoric Model:} This method only quantifies aleatoric uncertainty by using the Gaussian likelihood of the proposed framework with the maximum likelihood estimate of the parameters $\theta$. For this method, we have used the neural networks $f$ and $\sigma^2$ depicted in Figure \ref{fig:highlevel}, Figure \ref{fig:residual}, and Figure \ref{fig:unet}. The batch size was set to $4$ for the MRI experiments and $2$ for the CT experiments. Learning rate was fixed to $1\times 10^{-4}$ for the MRI experiments and $1\times 10^{-5}$ for the CT experiments. Initial step size of the neural network $f$ was set to $1.0$ for the MRI experiments and $1\times 10^{-4}$ for the CT experiments. The number of iterations of the PGD was set to be $5$, and the neural networks $f$ and $\sigma^2$ were trained for $100$ epochs.

\textbf{Proposed Only Epistemic Model:} This is the preliminary variant of the proposed method that we have presented in the prior conference publication~\cite{Ekmekci2021UncertaintyUnfoldingPreliminary}. This method only quantifies the epistemic uncertainty since it treats the covariance matrix of the Gaussian likelihood function as a fixed parameter. In our experiments, we fixed the covariance matrix to $(1/10) \mathbf{I}$. The batch size was set to $4$ for the MRI experiments and $2$ for the CT experiments. Learning rate was set to be $1\times 10^{-4}$ for the MRI experiments and $1\times 10^{-5}$ for the CT experiments. Initial step size of the neural network $f$ was fixed to $1.0$ for the MRI experiments and $1\times 10^{-4}$ for the CT experiments. The number of iterations of the PGD was set to be $5$, and the dropout-added neural network was trained for $100$ epochs.

\section{MC Dropout Primer}
For the sake of completeness, we state the assumptions of MC Dropout explicitly and discuss the variational inference steps. For a more detailed discussion, the reader can refer to ~\cite{Gal2016MCDropout, Gal2016BayesianCNN, Kendall2017BayesianNN}. Suppose that we have access to a training dataset $\mathcal{D} = \{ (\m^{[n]}, \s^{[n]}) \}_{n=1}^{N_\mathcal{D}}$ containing $N_\mathcal{D}$ pairs of measurement vectors and target images. Moreover, we assume that the neural networks $f_\gamma$ and $\sigma_\delta^2$ contain $C$ and $E$ convolutional layers, respectively. Then, we can write the two sets $\gamma$ and $\delta$ as follows:
\begin{equation}
\gamma =  \bigcup_{i=1}^{C} \{ \W_{i}^f \} \quad \text{and} \quad \delta =  \bigcup_{j=1}^E \{ \W_{j}^\sigma \},
\end{equation}
where $\W_{i}^f$ and $\W_j^\sigma$ are the matrices whose rows contain the vectorized filter coefficients of the $i^{\text{th}}$ and $j^{\text{th}}$ convolutional layers of the neural networks $f_\gamma$ and $\sigma^2_{\delta}$, respectively. The assumptions~\cite{Gal2016MCDropout, Gal2016BayesianCNN, Kendall2017BayesianNN} on the parametric distribution $q_\omega (\theta)$ that we use to approximate the true posterior distribution $p(\theta | \mathcal{D})$ are as follows: (i) For the parametric distribution, we assume that the layers of the neural networks $f_\gamma$ and $\sigma_\delta^2$ are independent, and layers within the neural networks are mutually independent, i.e.,
\begin{equation}
q_\omega(\theta) = \left( \prod_{i=1}^{C} q\left(\W_{i}^f\right) \right) \left( \prod_{j=1}^{E} q\left(\W_{j}^\sigma \right) \right);
\end{equation}
(ii) Filters of a convolutional layer are assumed to be mutually independent, more explicitly
\begin{equation}
\begin{aligned}
q(\W_{i}^f) = \prod_{l=1}^{K_{i, f}^{[out]}} q( [ \W_{i}^f ]_{l,:} ), 
q\left(\W_{j}^\sigma \right) = \prod_{m=1}^{K_{j,\sigma}^{[out]}} q( \left[ \W_{j}^\sigma \right]_{m,:} ),
\end{aligned}
\end{equation}
where $K_{i, f}^{[out]}$ is the number of filters in the $i^\text{th}$ convolutional layer of $f_\gamma$, and $K_{j,\sigma}^{[out]}$ is the number of filters in the $j^\text{th}$ convolutional layer of $\sigma^2_\delta$; (iii) The distribution of the filter coefficients of each filter is a mixture of Gaussians distribution defined by
\begin{equation}
\begin{aligned}
q( [ \W_{i}^f ]_{l,:} ) &= p(z_{i,l}^f=1) \mathcal{N}( [ \W_{i}^f ]_{l,:} | \a_{i,l}^f, \epsilon^2\I) \\
&\quad+p(z_{i,l}^f=0) \mathcal{N}( [ \W_{i}^f ]_{l,:} | \0, \epsilon^2\I), \\
q( [ \W_{j}^\sigma ]_{m,:} ) &= p(z_{j,m}^\sigma=1) \mathcal{N}([ \W_{j}^\sigma ]_{m,:} | \a_{j,m}^\sigma, \epsilon^2\I) \\
&\quad+ p(z_{j,m}^\sigma=0) \mathcal{N}([ \W_{j}^\sigma ]_{m,:} | \0, \epsilon^2\I),
\end{aligned}
\label{eq:bernoullivariationaldistribution}
\end{equation}
where the variables $z_{i,l}^f$ and $z_{j,m}^\sigma$ are the latent variables, and the scalars $p_{i,l}^f \triangleq p(z_{i,l}^f=1)$ and $p_{j,m}^\sigma \triangleq p(z_{j,m}^\sigma=1)$ are fixed constants. The scalar $\epsilon$ is a very small fixed constant, and the sets $\Delta_f \triangleq \{ \a_{i,l}^f \}$ and $\Delta_\sigma \triangleq \{ \a_{j,m}^\sigma \}$ are the adjustable parameters of the parametric distribution. Previously we have denoted the set of adjustable parameters of the parametric distribution $q_\omega(\theta)$ with $\omega$, so we can write the set $\omega$ explicitly as $\omega = \Delta_f \cup \Delta_\sigma$.

Under these assumptions, we may want to adjust the parameters of the parametric distribution by minimizing the Kullback-Leibler divergence between the parametric distribution and the true posterior distribution, i.e.,
\begin{equation}
\hat{\omega} = \argmin_\omega D_{\text{KL}} \left( q_\omega(\theta) || p(\theta|\mathcal{D}) \right).
\label{eq:klminimization}
\end{equation}
Unfortunately, directly solving the optimization problem in \eqref{eq:klminimization} requires the knowledge of the true posterior distribution of the parameters $p(\theta|\mathcal{D})$, which is what we do not know, and which is what we aim to obtain an estimate of. An alternative to minimizing the KL divergence is to maximize the log-evidence lower bound, which leads to the following optimization problem.
\begin{equation}
\hat{\omega} = \argmax_\omega \left\{ \mathbb{E}_{\theta \sim q_\omega(\theta)} \left[ \log p(\mathcal{S}| \mathcal{M}, \theta )\right] - D_{\text{KL}} \left( q_\omega(\theta) || p(\theta) \right) \right\},
\label{eq:logevidencelowerboundmaximization}
\end{equation}
where $\mathcal{S}= \{ \s^{[n]} \}_{n=1}^{N_\mathcal{D}}$ is the set containing the training target images, and $\mathcal{M}= \{ \m^{[n]} \}_{n=1}^{N_\mathcal{D}}$ is the set containing the training measurements. Assuming that the training dataset $\mathcal{D}$ contains i.\ i.\ d.\ examples, MC Dropout approximates the expectation term in \eqref{eq:logevidencelowerboundmaximization} as follows:
\begin{equation}
\begin{aligned}
&\mathbb{E}_{\theta \sim q_\omega(\theta)} \left[ \log p(\mathcal{S}| \mathcal{M}, \theta )\right] = \int q_\omega(\theta) \log p(\mathcal{S} | \mathcal{M}, \theta) d \theta \\
&= \int q_\omega(\theta) \left( \sum_{n=1}^{N_\mathcal{D}} \log p(\s^{[n]} | \m^{[n]}, \theta) \right) d \theta \\
&=  \sum_{n=1}^{N_\mathcal{D}}  \int q_\omega(\theta) \log p(\s^{[n]} | \m^{[n]}, \theta) d \theta \\
&\approx  \sum_{n=1}^{N_\mathcal{D}}  \log p(\s^{[n]} | \m^{[n]}, \tilde{\theta}^{(n)}) d \theta, \qquad \tilde{\theta}^{(n)} \sim q_\omega(\theta) \\
&= \sum_{n=1}^{N_\mathcal{D}}  \log  \mathcal{N}(\s^{[n]}| f_{\tilde{\gamma}^{(n)}}(\m^{[n]}), \diag(\sigma_{\tilde{\delta}^{(n)}}^2(\m^{[n]}))) d \theta \\
&= \sum_{n=1}^{N_\mathcal{D}} \sum_{k=1}^N \bigg[ -\frac{1}{2} \log 2 \pi - \frac{1}{2} \log  [\sigma_{\tilde{\delta}^{[n]}}^2(\m^{[n]})]_k \\
&\qquad \qquad \qquad \qquad -\frac{( [\s^{[n]}]_k - [f_{\tilde{\gamma}^{(n)}} (\m^{[n]})]_k )^2}{2 \sigma_{\tilde{\delta}^{[n]}}^2(\m^{[n]})]_k} \bigg],
\label{eq:expectationterm}
\end{aligned}
\end{equation}
where the second line follows from the independence assumption, and the fourth line follows from approximating the integral in the third line using Monte Carlo integration with one sample. 

To calculate the second term of the objective function of the optimization problem in \eqref{eq:logevidencelowerboundmaximization}, we need to compute the KL divergence between the candidate distribution and the prior distribution of the parameters. Assuming that the prior distribution on the parameters have the form of a standard Gaussian distribution for each parameter in the set $\theta$, Gal and Ghahramani~\cite{Gal2016MCDropout, Gal2016BayesianCNN} approximate KL divergence term as follows:
\begin{equation}
\begin{aligned}
&D_{\text{KL}} \left( q_\omega(\theta) || p(\theta) \right) \approx \sum_{i=1}^{C} \sum_{l=1}^{K_{i,f}^{[out]}} C^f_{i,l} +\frac{p_{i,l}^f}{2} \| \a_{i,l}^f \|_2^2 \\
&\qquad \qquad \qquad \qquad + \sum_{j=1}^E \sum_{m=1}^{K_{j, \sigma}^{[out]}} C^\sigma_{j,m} + \frac{p_{j,m}^\sigma}{2} \| \a_{j,m}^\sigma \|_2^2,
\label{eq:klterm}
\end{aligned}
\end{equation}
where $C^f_{i,l}$ and $C^\sigma_{j,m}$ are constants that do not depend on the set $\omega$ of parameters.

Using these two expressions, the optimization problem in \eqref{eq:logevidencelowerboundmaximization} can be written as follows.
\begin{equation}
\begin{aligned}
\hat{\omega} &\approx \argmin_\omega \bigg\{ \frac{1}{N_\mathcal{D}} \sum_{n=1}^{N_\mathcal{D}} \sum_{k=1}^{N} \bigg[ \frac{1}{2} \log [\sigma_{\tilde{\delta}^{(n)}}^2 (\m^{[n]})]_k \\
&+ \frac{1}{2} \exp( -  \log [\sigma_{\tilde{\delta}^{(n)}}^2 (\m^{[n]})]_k) ( [\s^{[n]}]_k - [f_{\tilde{\gamma}^{(n)}} (\m^{[n]})]_k )^2 \bigg] \\
&+\sum_{i=1}^{C} \sum_{l=1}^{K_{i,f}^{[out]}} \frac{p_{i,l}^f}{2 N_\mathcal{D}} \| \a_{i,l}^f \|_2^2 + \sum_{j=1}^E \sum_{m=1}^{K_{j, \sigma}^{[out]}} \frac{p_{j,m}^\sigma}{2 N_\mathcal{D}} \| \a_{j,m}^\sigma \|_2^2
\bigg\}.
\end{aligned}
\end{equation}

\section{Comparison with the Uncertainty Quantifying Plug-and-Play Method}

Ekmekci and Cetin~\cite{Ekmekci2021UncertaintyPnP} have previously proposed a Plug-and-Play (PnP) method that aims to quantify the epistemic uncertainty of the parameters of the CNN denoiser used within the update equations of the PnP algorithm. In this section, we compare the PnP method presented in \cite{Ekmekci2021UncertaintyPnP} and the deep unrolling-based method presented in this work. 

One of the main conceptual differences between the PnP method in \cite{Ekmekci2021UncertaintyPnP} and the proposed method is that the PnP method is not an end-to-end model, whereas the proposed method is an end-to-end model. Also, the PnP method presented in \cite{Ekmekci2021UncertaintyPnP} only quantifies the epistemic uncertainty on the parameters of the CNN denoiser and does not focus on the aleatoric uncertainty. On the other hand, the method presented in this work aims to quantify the aleatoric uncertainty as well by modeling the diagonal entries of the covariance matrix of the Gaussian likelihood function with a neural network. The main advantage of the PnP method is that it is more modular in the sense that it does not require training for different configurations of the imaging setup. This makes the PnP method more desirable for applications where the configurations of the imaging setup may change frequently. 

\begin{figure}[!t]
    \centering
    \includegraphics[width=0.9\columnwidth]{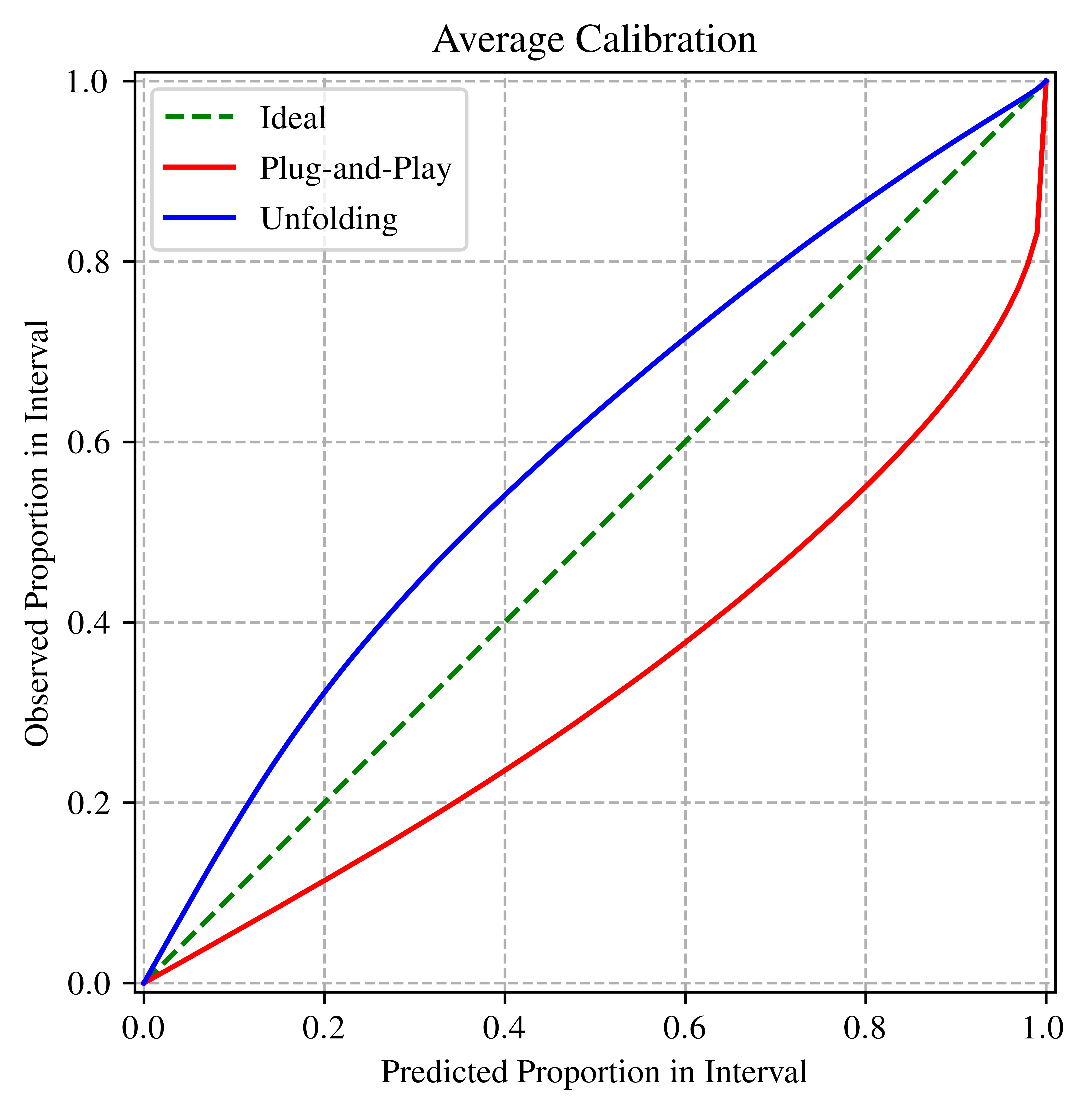}
    \caption{Calibration plots of the proposed method and the PnP method for an MRI experiment, where the undersampling rate is $20\%$ and SNR is 70 dB.}
    \label{fig:calibration_plots_pnp}
\end{figure}

One quantitative way to compare the quality of the predictive uncertainty estimates of the PnP method and the proposed framework is to measure the calibration and sharpness of the predictive distributions provided by these methods. For this purpose, we tested the PnP method and the proposed method on an MRI simulation using the test dataset, where the undersampling rate was set to $20\%$, and the SNR was fixed to 70 dB. Then, we generated the calibration plot and calculated calibration metrics such as mean-absolute calibration error (MACE), root-mean-squared calibration error (RMSCE), and miscalibration area (MA), and the sharpness of these methods using the Uncertainty Toolbox~\cite{chung2021uncertainty}. Table \ref{tab:calibrationandsharpnesspnpandproposed} shows the values of the metrics, and Figure \ref{fig:calibration_plots_pnp} depicts the calibration plots of these methods. As can be seen from the Table \ref{tab:calibrationandsharpnesspnpandproposed}, the proposed method achieves lower calibration errors compared to the PnP method and provides as sharp probabilistic predictions as the PnP method. Figure \ref{fig:calibration_plots_pnp} shows that the proposed method provides slightly underconfident probabilistic predictions while the PnP method may provide overconfident probabilistic predictions. 



\begin{table}[!t]
\centering
\caption{Calibration and sharpness metrics of the PnP method and the proposed method calculated over the MRI test dataset, where the percentage of the observed k-space coefficients is $20 \%$, and SNR is $70$ dB.}
\label{tab:calibrationandsharpnesspnpandproposed}
\begin{tabular}{l|cc}
\toprule
Metrics & PnP Method & Proposed Method \\
\midrule
MACE & 0.1640 & \textbf{0.0917} \\
RMSCE & 0.1817 & \textbf{0.1018} \\
MA & 0.1657 & \textbf{0.0926} \\
\midrule
Sharpness & \textbf{0.0232} & 0.0235 \\
\bottomrule
\end{tabular}
\end{table}

\section{Aleatoric Uncertainty and Reducibility}

\begin{figure}[!t]
    \centering
    \includegraphics[width=0.9\columnwidth]{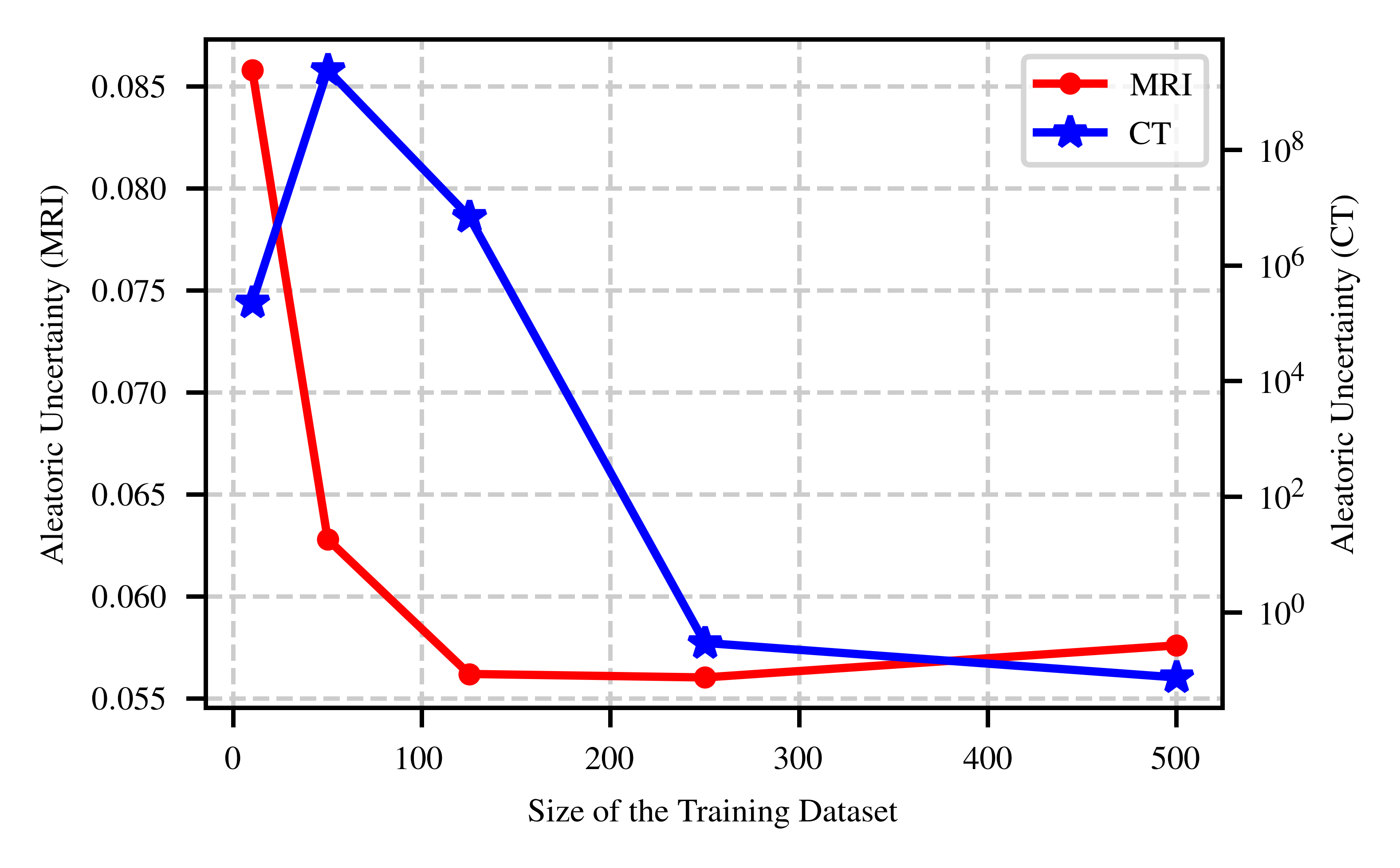}
    \caption{Mean aleatoric uncertainty as a function of training dataset size. For the MRI experiments, the percentage of observed k-space coefficients is $20\%$, and SNR is 70 dB. For the CT experiments, number of views is $36$, and SNR is 70 dB. We experimentally observe that the average aleatoric uncertainty becomes almost constant after having certain number of training examples.}
    \label{fig:aleatoric_reducibility}
\end{figure}

By definition, aleatoric uncertainty is the inherent uncertainty in the reconstruction task caused by the ill-posed nature of an imaging inverse problem; therefore, unlike epistemic uncertainty, it is not affected by the number of training samples and is not reducible. On the other hand, the proposed method models the aleatoric uncertainty by representing the diagonal entries of the covariance function of the likelihood function with a U-shaped neural network, which is a parametrized function. Thus, we also have uncertainty on the parameters of the U-shaped function. In the Bayesian neural network formulation, this uncertainty is taken into account while constructing the aleatoric uncertainty maps by averaging the output of the U-shaped neural network over the possible values of the parameters of the U-shaped neural network. Hence, the resulting aleatoric uncertainty information provided by the proposed framework depends on the size of the training dataset since a small training dataset may not be enough to reduce the epistemic uncertainty on the parameters of the U-shaped neural network. Figure \ref{fig:aleatoric_reducibility} shows the average aleatoric uncertainty as a function of the size of the training dataset for CT and MRI reconstruction problems. We experimentally observe that the average aleatoric uncertainty becomes almost constant after having certain number of training examples, which may be sufficient to reduce the epistemic uncertainty on the parameters of the U-shaped network. Based on this observation, to achieve reliable aleatoric uncertainty estimates, at the training stage of the proposed method, different subsets of the training dataset having different sizes can be used to train the proposed model. Then, the resulting models can be tested on a validation dataset to observe how epistemic and aleatoric uncertainties change as a function of the size of the training datasets, e.g., one can compute average epistemic and aleatoric uncertainties over all pixels. If the aleatoric uncertainty estimates provided by the proposed method becomes almost constant at some point, we can choose that point as the point where the epistemic uncertainty on the parameters of the U-shaped network has decreased sufficiently.

\section{Reconstruction Performance}
\begin{figure*}[t]
	\centering
	\includegraphics[width=0.95\textwidth]{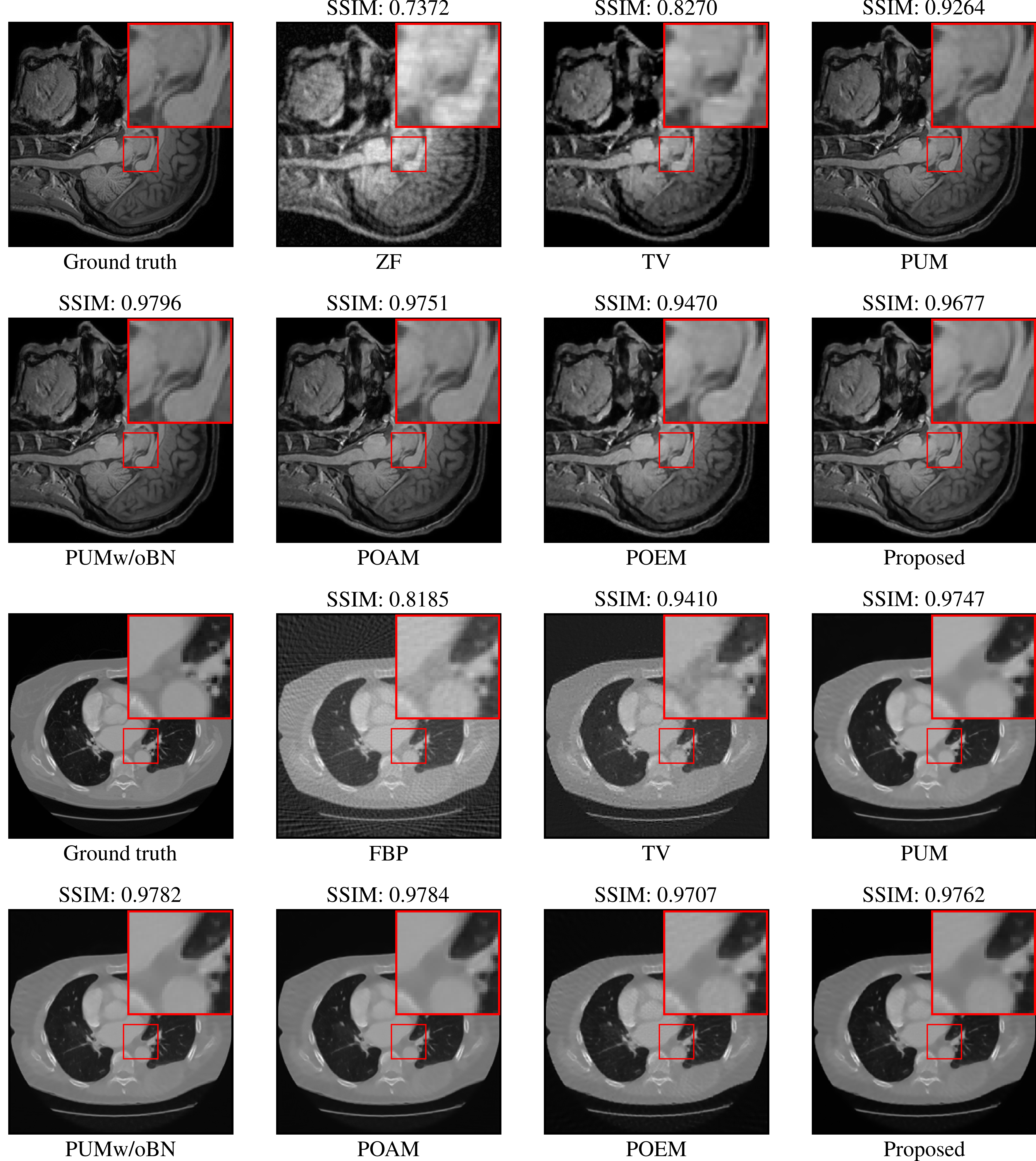}
	\caption{Visual comparison of the image reconstruction performance of zero-filling (ZF) / filtered backprojection (FBP), total variation reconstruction (TV), PGD-based deep unrolling method (PUM), PGD-based deep unrolling method without batch normalization (PUMw/oBN), proposed only epistemic model (POEM), proposed only aleatoric model (POAM), and the proposed method.}
	\label{fig:reconstructionexample}
\end{figure*}

Figure \ref{fig:reconstructionexample} compares the reconstruction performance of the proposed method with the reconstruction methods whose details are discussed in Section \ref{sec:othermethods} of the Supplementary Material.

\bibliographystyle{IEEEtran}
\bibliography{refs_supplementary}